\documentclass[12pt]{JHEP3}
\usepackage{epsfig}
\usepackage{amsmath}
\def\beq{\begin{eqnarray}}
\def\eeq{\end{eqnarray}}
\title{
Solitons in Supersymmety Breaking Meta-Stable Vacua}
\author{
Minoru Eto$^{\dagger a}$, 
Koji Hashimoto$^{\dagger b}$ and 
Seiji Terashima$^{* c}$\\
${}^\dagger$ {\it Institute of Physics, the University of Tokyo, Komaba,
Tokyo 153-8902, Japan}\\
${}^*$ {\it Yukawa Institute for Theoretical Physics, \\
\hspace{40mm}Kyoto University, Kyoto 606-8502, Japan}\\
$^a$ E-mail: \email{meto@hep1.c.u-tokyo.ac.jp}\\
$^b$ E-mail: \email{koji@hep1.c.u-tokyo.ac.jp}\\
$^c$ E-mail: \email{terasima@yukawa.kyoto-u.ac.jp}\\
}

\abstract{
In recently found supersymmetry-breaking meta-stable vacua of the 
supersymmetric
QCD, we examine possible exsitence of solitons. Homotopy groups of the
moduli space of the meta-stable vacua show that there is no nontrivial
soliton for $SU(N_c)$ gauge group. 
When $U(1)_B$ symmetry present in the theory is gauged, we find
non-BPS solitonic (vortex) strings whose existence and properties are 
predicted from brane configurations. We obtain explicit classical
solutions which reproduce the predicitions. 
For $SO(N_c)$ gauge group, 
we find there are solitonic strings for $N \equiv N_f-N_c+4=2$, 
and ${\bf Z}_2$ strings for the other $N$.  
The strings are meta-stable 
as they live in the meta-stable vacua.
}

\preprint{
{\normalsize{\tt hep-th/0610042}}\\
{\normalsize UT-Komaba/06-10}\\
{\normalsize YITP-06-50}
}

\begin{document}

\section{Introduction}
\label{section1}

There are many ways to motivate need of supersymmetries --- 
while one is, needless to say, a solution to the loop-level hierarchy 
problem, others include a strong controllability of the
theories. Renowned Seiberg's duality \cite{Seiberg:1994pq}
among ${\cal N}=1$ supersymmetric 
gauge theories is a famous example. Furthermore, with the
supersymmetries, analysis via string/M theories are accessible from 
field theories. However, the introduction of the supersymmetries in realistic 
field theory models always accompanies the problem of how to break them
at low energy. In this sense, it should be celebrated that
Intriligator, Seiberg and Shih \cite{ISS} found recently that 
one of the simplest supersymmetric field theories, the supersymmetric
QCD (SQCD),  
admits supersymmetry-breaking meta-stable vacua via the Seiberg's
duality (some following works include 
\cite{OO2,Franco:2006ht,Bena:2006rg,Franco:2006es,follow}). 
With the help of 
known realization of the duality in string theories
\cite{Elitzur:1997fh} a la Hanany-Witten setup \cite{Hanany:1996ie} 
(see \cite{Giveon:1998sr} for a review), the
brane configuration corresponding to the 
meta-stable vacua have been identified
\cite{OO2,Franco:2006ht,Bena:2006rg}, even though the vacua 
break the supersymmetries. 

Since this provides a new mechanism to break the supersymmetries, 
it is important to study classical/quantum properties of this newly
found vacua. In particular, the structure of the classical moduli space 
is directly related to a possible existence of solitons in the vacua.
Any soliton, if existent, is quite relevant to particle phenomenology, 
partially through cosmological evolution and phase transition of
the universe.

In this article, we examine the possible existence of solitons, in the
supersymmetry-breaking meta-stable vacua. Since the vacua are provided
by a Seiberg-dual (called ``magnetic'' or ``macroscopic'') theory of the
SQCD, field-theoretical facilitation is concrete enough to determine
topology of the moduli space. We find that when the gauge group of the
SQCD is $SU(N_{\rm c})$, homotopy groups of the moduli space are trivial,
thus any soliton does not appear in the vacua. 

When the gauge group is $U(N_{\rm c})$, mainly because of its diagonal
$U(1)$ sector, the vacua admit vortex strings. With a help of the
brane realization of the meta-stable vacua 
\cite{OO2,Franco:2006ht,Bena:2006rg} (see also \cite{Franco:2006es}), 
the existence of the vortex strings is predicted from brane
configurations, as a generalization of  
Hanany-Tong set-up \cite{Hanany:2004ea} where vortex strings in 
${\cal N}=2$ gauge 
theories are identified with D2-branes in Hanany-Witten brane
configurations. The predicted properties such as tensions,
supersymmetries, and species of strings, 
can be examined by direct construction of
vortex string solutions in the magnetic theory of the SQCD. 
It is noteworthy that the vortex strings are non-BPS ($\equiv$
supersymmetry-breaking). Eventually 
the magnetic theory is found to be a 
gauge theory with flavor fields allowing so-called semilocal vortex
strings \cite{Vachaspati:1991dz}, which are our non-BPS vortex strings.
Because they are topological,
the lifetime of the corresponding strings in the SQCD 
is of the same order as the meta-stable vacua.
When the quark masses of the original SQCD split, various kinds
of vortex strings appear with different tensions.

For $SO$ gauge groups, we find similar semilocal strings only
when $N_c=N_f+2$ where the dual gauge group is $SO(2)\sim U(1)$.
For generic $N_f$ and $N_c$ we find
${\bf Z}_2$ solitonic strings.

The organization of the paper is as follows. After reviewing
the supersymmetry-breaking meta-stable vacua of the magnetic SQCD
\cite{ISS}, we
examine the homotopy groups of the moduli space of the vacua in section
\ref{sec:soliton}. Possible existence of solitons is studied
accordingly. Then in section \ref{sec:branestring}, we obtain a brane
configuration of the vortex strings in 
the meta-stable vacua, following the 
brane realization of the meta-stable vacua
\cite{OO2,Franco:2006ht,Bena:2006rg} and also that of the solitonic
strings in ${\cal N}=2$ gauge theories \cite{Hanany:2004ea}.
In section \ref{sec:solution}, we explicitly construct the vortex string
solutions  
in the magnetic SQCD, and study their stability to see the consistency
with the brane picture. Some discussions on renormalization groups,
relevance to cosmologies, and detailed correspondence to the brane
picture are presented in section 
\ref{sec:conclusion}, with a brief conclusion.

\section{Solitons in Magnetic Theory of SQCD}
\label{sec:soliton}
\setcounter{footnote}{1}

\subsection{Meta-stable vacua of SQCD}
\label{sec:iss}

First, we briefly summarize 
the main results of Intriligator, Seiberg and Shih \cite{ISS}, 
the supersymmetry-breaking meta-stable vacua of the magnetic
theory of the SQCD. 

In \cite{ISS},
${\cal N}=1$ $4d$ supersymmetric $SU(N_c)$ Yang-Mills with 
$N_f$ massive flavors $Q$ and $\widetilde Q$ (SQCD) 
was shown to have meta-stable supersymmetry-breaking 
vacua 
in the free magnetic range $N_c+1 \le N_f \le \frac{3}{2}N_c$,
with sufficiently small quark masses.
Along the analysis done in \cite{ISS}, we will work in
this range.
The dual $SU(N)$ $(N=N_f-N_c)$ theory is infra-red free and its low
energy effective theory is described by the K\"ahler potential and the
superpotential 
\beq
K &=& {\rm Tr}_c
\left[q e^{-V} q^\dagger 
+ \widetilde q^\dagger e^V \widetilde q \right]
+ {\rm Tr}_f \left[M^\dagger M \right],\\
W &=& h {\rm Tr}_c \left[q M \widetilde q\right] 
- h \mu^2 {\rm Tr}_f M.
\label{superpotential}
\eeq
Here $M$ is the meson field and $q,\widetilde q$ are the dual quarks
whose charges are summarized below.
(We omit the $U(1)_R$ charge in the table.)
\begin{center}
\begin{tabular}{c|ccc}
& $SU(N)$ & $SU(N_f)_V$ & $U(1)_B$ \\
\hline\hline
$M_{[N_f \times N_f]}$ & 1 & adj. & $0$ \\
$q_{[N \times N_f]}$ & $\square$ & $\bar \square$ & $1$\\
$\widetilde q_{[N_f \times N]}$ & $\bar \square$ & $\square$ & $-1$\\
\end{tabular}
\end{center}
\ 

The bosonic part of the 
Lagrangian of the ``macroscopic theory'' which we call magnetic theory
is of the form 
\beq
{\cal L} = {\rm Tr}_c
\left[ - \frac{1}{2g^2} F_{\mu\nu} F^{\mu\nu}
- {\cal D}_\mu q {\cal D}^\mu q^\dagger
- {\cal D}_\mu \widetilde q^\dagger {\cal D}^\mu \widetilde q
\right]
- {\rm Tr}_f \left[
\partial_\mu M^\dagger \partial^\mu M
\right]
- V
\eeq
with a scalar potential $V = V_F + V_D$ given by\footnote{
When the gauge group is $U(N)$ instead of the $SU(N)$ ({\it i.e.} when
we gauge the $U(1)_B$ with its coupling put equal to that of the
$SU(N)$), the second term in $V_D$ is not necessary.}
\beq
V_F &=& |h^2|
{\rm Tr}_f \left[
\left|\widetilde q q - \mu^2 {\bf 1}_{N_f} \right|^2
\right] +\ |h^2| {\rm Tr}_c \left[
\left| q M \right|^2
+ \left| \widetilde q^\dagger M^\dagger \right|^2
\right],
\label{iss_pot_1}\\
V_D &=& \frac{g^2}{4} {\rm Tr}_c
\left[
\left(q q^\dagger - \widetilde q^\dagger \widetilde q\right)^2
\right]
-\frac{g^2}{8} 
\left({\rm Tr}_c q q^\dagger - 
{\rm Tr}_c\widetilde q^\dagger \widetilde q\right)^2.
\label{iss_pot_2}
\eeq
Supersymmetric configuration 
is then given by
\beq
\widetilde q q = \mu^2 {\bf 1}_{N_f},\quad
q M = 0,\quad
\widetilde q^\dagger M^\dagger = 0,\quad
q q^\dagger - \widetilde q^\dagger \widetilde q = 0.
\label{vacuum}
\eeq
The first condition cannot be satisfied when 
$N < N_f$ because the rank $N$ of the matrix $\widetilde q q$ 
is less than $N_f$ (rank condition) \cite{ISS}.
A configuration minimizing the potential
is of the form 
\beq
M =
\left(
\begin{array}{cc}
0_{N \times N} & 0_{N \times (N_f-N)}\\
0_{(N_f-N) \times N} & M_{0}
\end{array}
\right),\quad
\frac{q^\dagger}{\mu^*} = \frac{\widetilde q }{\mu}
=  \left(
\begin{array}{c}
{\bf 1}_N\\
0_{(N_f-N) \times N}
\end{array}
\right),
\label{VEV}
\eeq
where $M_0$ is an arbitrary $N_f - N$ by $N_f - N$ matrix.
Except for the first condition in equation (\ref{vacuum}),
all the other conditions are satisfied. So the vacuum energy is given by
\beq
V = |h \mu^2|^2 (N_f - N) > 0,
\label{meta}
\eeq
and thus the vacuum spontaneously breaks the ${\cal N}=1$ supersymmetry.
We will see this supersymmetry breaking from the view
point of the D-brane configuration in section \ref{sec:issbrane}.

Notice that this vacuum shows color-flavor locking which
is invariant under the $SU(N)_{c+f}$ global transformation,
\beq
M \to UM U^\dagger = M,\quad
q \to g q U^\dagger = q,\quad
\widetilde q \to U \widetilde q g^\dagger = \widetilde q.
\eeq
Here $M,q,\widetilde q$ are given in (\ref{VEV}), 
$g \in SU(N)$ is a global rotation of the gauge group $SU(N)$, 
and $U$ is an element of $SU(N_f)$ defined by
\beq
U = 
\left(
\begin{array}{cc}
g & \\
& {\bf 1}_{N_f - N}
\end{array}
\right) \in SU(N_f).
\eeq

The supersymmetry-breaking vacua given in (\ref{VEV})
have several flat directions. A part of them
are truly massless Nambu-Goldstone 
modes associated with the spontaneously broken
global symmetries, while the others are classical pseudo-moduli fields
which acquires positive masses by one-loop contributions 
to the effective potential \cite{ISS}.
The vacuum of the potential is then $M_0 = 0$ and 
the Nambu-Goldstone modes are generated by
\beq
q = 
\left( \mu {\bf 1}_N,\ 0 \right) \to 
\hat g \left( \mu {\bf 1}_N,\ 0 \right) \hat U^\dagger,
\eeq
where $\hat g \in SU(N)$ and $\hat U \in U(N_f)
\sim SU(N_f) \times U(1)_B$.
We will study in detail the vacuum manifold of this magnetic theory in
the following subsections.

In the SQCD above, the masses of the quarks are chosen to be the
same, which are proportional to $\mu^2$. 
We can introduce different mass for each quark, by
replacing the superpotential (\ref{superpotential}) by
\begin{eqnarray}
 W &=& h {\rm Tr}_c \left[q M \widetilde q\right] 
- h {\rm Tr}_f \left[mM\right],
\end{eqnarray}
where $m={\rm diag} (m_1,m_2, \cdots, m_{N_f})$, and $m_i$ is the quark
mass of the SQCD times a dynamical scale $(-\hat{\Lambda})$, see
\cite{ISS}. 
We can bring $m_i$ real and positive classically, and 
choose $m_1\geq m_2\geq \cdots \geq m_{N_f}$.
When all the masses are
non-vanishing, the supersymmetry is broken at the vacuum. However, when
$m_{N+1}=\cdots = m_{N_f}=0$, the rank condition is satisfied, and the
supersymmetry is unbroken at the vacuum although
the vacuum expectation value is the same as that given in equation 
(\ref{VEV}).
Notice that,
by the mass arrangement 
$m = {\rm diag}(\underbrace{\mu^2,\cdots,\mu^2}_N,
\underbrace{0,\cdots,0}_{N_f - N})$, 
the flavor symmetry $SU(N_f)$ is explicitly broken down to
$SU(N)$.  

\subsection{Topological solitons}
\label{sec:topsol}

Let us study the possible existence of 
the topological
solitons in the meta-stable vacua in the magnetic theory.
If there are topological solitons,
the corresponding solitons in the meta-stable vacua in the SQCD 
have lifetime of the order of the lifetime of the meta-stable vacua.

We also  study the moduli space ${\cal M}_{\rm vac}$
of the meta-stable vacua and the topological solitons
in the nonlinear sigma model whose target spaces is
the moduli space. 
These nonlinear sigma models are obtained 
as the effective action for the energy lower than 
the mass scale of the pseudo Nambu-Goldstone modes.
It is important to note that
there is no obstruction to continuously deform
the configuration 
corresponding to the nonlinear sigma model soliton
to the meta-stable vacua, in the magnetic theory.
Moreover, this deformation is not 
related to the true supersymmetric vacua and 
is closed in the region near the meta-stable vacua.
Thus
the nonlinear sigma model solitons 
have a relatively short lifetime
of the order of the scale given by the 
mass of the pseudo Nambu-Goldstone modes
(if they do not correspond to any topological soliton
of the magnetic theory of the SQCD).

In advance, let us list our results here. 
In the magnetic theory of 
the SQCD with the gauge group $SU(N_c)$,
the gauge group $SU(N)$ is completely broken 
at the meta-stable vacuum.
Since $\pi_k(SU(N))=0$ for $0 \leq k \leq 2$,
we expect there is no topological soliton
except (constrained) instantons in its Euclideanized theory.
In fact, we find the homotopy groups of
the moduli space of vacua, as  
\begin{eqnarray}
 \pi_0({\cal M}_{\rm vac})=
 \pi_1({\cal M}_{\rm vac})=
\pi_2({\cal M}_{\rm vac})=0.
\label{hom1}
\end{eqnarray}
Thus we expect no topological solitons (global monopole 
/ vortex string / domain wall) in the meta-stable vacua, and
there is no string or domain wall
even in the sigma model which is the low energy effective theory\footnote{
In this paper 
we do not consider the particle like solitons in the sigma model
although $\pi_3({\cal M}_{\rm vac})$ is not always trivial.
The reasons are as follows.
The Derrick's theorem
states that there are no stable sigma model solitons whose co-dimension
is more than 1. The theorem is valid with no
higher derivative terms, so a smooth sigma model soliton 
might exist if we include higher derivative corrections, like skyrmions.
However, this corresponds to going to higher energy regime beyond
the non-linear sigma model limit. We do not have any natural reason why
one can abandon pesudo moduli and higher exited modes 
while keeping the paticular higher derivatgive terms on the moduli fields.
Even when one assumes such truncation, the lifetime of the sigma model
soliton is quite short compared to the lifetime of the
meta-stable vacua. 
We would like to thank K.~Intriligator on this point.
}.

In order to have non-trivial solitons, we need to consider  
$U(N_c)$ gauge group instead of the $SU(N_c)$, in other words,
gauging the $U(1)_B$ symmetry. 
This is equivalent to considering
$SU(N_c)\times U(1)_B/ {\bf Z}_N$ 
gauge group in the magnetic theory.\footnote{
Since the gauge couplings of the $SU(N_c)$ and the $U(1)_B$ are 
generically
different, the $SU(N_c)\times U(1)_B/ {\bf Z}_N$ is not 
equivalnet to $U(N)$ although they are topologicaly equivalent.
Note that ${\bf Z_N} \in SU(N)$ and ${\bf Z_N} \in U(1)_B$ 
act in the same way on the matter fields in the magnetic theory.}
In this case, 
the gauge group $SU(N) \times U(1)_B$ is completely broken 
at the meta-stable vacuum and then
we expect to have (semi-)local vortex strings from the broken $U(1)_B$.
This is the meta-stable string 
in the SQCD with the gauged $U(1)_B$. 
The moduli space of vacua 
$\widetilde{{\cal M}}_{\rm vac}$ has the following homotopy groups: 
\begin{eqnarray}
 \pi_0(\widetilde{{\cal M}}_{\rm vac})=
 \pi_1(\widetilde{{\cal M}}_{\rm vac})=0, \quad 
\pi_2(\widetilde{{\cal M}}_{\rm vac})={\bf Z}.
\label{hom2}
\end{eqnarray}
Therefore the gauged U(1) allows us to have global monopoles and sigma
model strings (lumps). The global monopoles are usually forbidden since
they have infinite mass. 
The sigma model strings correspond to the (semi-)local vortex strings
in the magnetic theory.

When the gauge group is $SO(N_{\rm c})$, the condition to have the
ultra-violet SQCD is $N_f<(3/2)(N_c-2)$. The gauge group of the dual
magnetic theory is $SO(N)$ with $N=N_f-N_c+4\geq 1$.
For the breaking of the gauge group, we have $\pi_1(SO(N))={\bf Z}_2$
except for the case of $SO(2)\sim U(1)$ where $\pi_1(SO(2))={\bf Z}$.
For the moduli space, we find 
\begin{eqnarray}
&&\pi_0 ({\cal M}_{SO} ) =0, \quad 
\pi_1 ({\cal M}_{SO} )=
\left\{
\begin{array}{ll}
{\bf Z}  & \mbox{(for $(N_f,N_c)=(2,5)$)}\\ 
0 & \mbox{(for the other cases)}
\end{array}
\right.
\nonumber \\
&&\pi_2 ({\cal M}_{SO} ) = 
\left\{
\begin{array}{ll}
{\bf Z}  & \mbox{(for $N_c=N_f+2$, except $N_f=4$)}\\ 
{\bf Z}_2 & \mbox{(for the other cases)}
\end{array}
\right.
\label{sohomo}
\end{eqnarray}
Thus, 
when $N=2$ (or equivalently $N_c=N_f+2$), the gauge group is 
$SO(2)\sim U(1)$ which is broken, thus
there are semilocal strings.\footnote{For the special case $N_f=4$, see
the next subsection.} 
For generic $N$ we find ${\bf Z}_2$ solitonic
strings.\footnote{For the special case of $(N_f,N_c)=(2,5)$, sigma-model
domain walls / global strings are possible.}
These strings are
topological solitons, hence are
meta-stable in the original SQCD.

In the rest of this section, first we shall obtain the moduli space of
the meta-stable vacua, with the un-gauged/gauged $U(1)_B$ symmetry for
the $SU(N_c)$ gauge groups. 
Then we will derive the homotopy groups (\ref{hom1}), (\ref{hom2})
and (\ref{sohomo}). 
The case with the $SO(N_c)$ will be described briefly.
In the next section, we study the brane realization of the (semi-)local
vortex strings for the case of the gauged $U(1)_B$, 
and in section \ref{sec:solution} we explicitly
construct classical solutions of the vortex strings.

\subsection{Moduli space of vacua}
\label{sec:msv}

The global symmetries of the magnetic theory of the SQCD are
$SU(N_f) \times U(1)_B$, where the $U(1)_B$ symmetry acts on the 
overall phase of the field $q$. The local symmetry is $SU(N)$.
We will deal with the case of the gauged $U(1)_B$ and $SO(N)$ gauge
group later.

The moduli space of the vacua is defined as a quotient space $G/H$, 
where $G$ is the global symmetry of the theory, and $H$ is global
symmetries which leave the vacuum invariant. Obviously in our case 
\begin{eqnarray}
 G = U(N_f).
\end{eqnarray}
Note that when we wrote the global symmetries as 
$SU(N_f) \times U(1)_B$,  we were not precise concerning 
the discrete subgroup,
and in fact $U(N_f)=(SU(N_f) \times U(1)_B)/{\bf Z}_{N_f}$ is the correct
global symmetry of the theory.\footnote{In the $U(N_f)$, elements 
$\exp[2\pi i n/N_f] {\bf 1}_{N_f}$ 
($n=0,1,2,\cdots,N_f-1$) belong to both the
$SU(N_f)$ and the $U(1)_B$. } Let us look for the group $H$. 
Consider the following elements of a subgroup 
$S\left(U(N) \times U(N_f-N)\right)$
of the $SU(N_f)$:
\beq
U =
\left(
\begin{array}{cc}
e^{i\frac{\theta}{N}}g_N & \\
& e^{-i\frac{\theta}{N_f-N}} g_{N_f-N}
\end{array}
\right)
\label{aho}
\eeq
where we have defined $g_M \in SU(M)$.
Consider the $S[U(1)\times U(1)]$ factor by choosing
$g_N = {\bf 1}_N$ and $g_{N_f-N} = {\bf 1}_{N_f-N}$,
\beq
\widetilde U = 
\left(
\begin{array}{cc}
e^{i\frac{\theta}{N}} {\bf 1}_N & \\
& e^{-i\frac{\theta}{N_f-N}} {\bf 1}_{N_f-N}
\end{array}
\right).
\label{aho2}
\eeq
We call this $\widetilde{U}(1)$ 
symmetry. The other $U(1)_B$ symmetry together with this 
$\widetilde{U}(1)$ act on the vacuum as
\beq
\mu \left({\bf 1}_N,\ 0 \right) \to
\mu e^{i\theta_B} \left({\bf 1}_N,\ 0 \right) \widetilde U^\dagger
=  \mu \left( e^{i\left(\theta_B - \frac{\theta}{N}\right)} {\bf 1}_N ,\ 0 \right).
\eeq
Therefore, when we have a relation
\begin{eqnarray}
 N\theta_B= \theta
\end{eqnarray}
the vacuum is invariant.
The combined symmetry is written as an element of $U(N_f)$,
\beq
U' = 
\left(
\begin{array}{cc}
{\bf 1}_N & \\
& e^{-i\frac{\theta}{N_f-N}} {\bf 1}_{N_f-N}
\end{array}
\right).
\label{aho3}
\eeq
We call this $U'(1)$ symmetry.

The element $g_N$ in the upper-left block in (\ref{aho}), 
when it acts on the vacuum, 
can be absorbed by the $SU(N)$ gauge symmetry which acts on $q$
from the left hand side. Thus, $SU(N)$ is still a symmetry of the
vacuum: this is the color-flavor locking. 
Furthermore, the element $g_{N_f-N}$ in the bottom-right block
in (\ref{aho}) is just an isotropy for the vacuum. Noting that 
together with the above $U'(1)$, this $SU(N_f-N)$ is upgraded to form
a $U(N_f-N)$, after considering the discrete subgroup $Z_{N_f-N}$
properly. Therefore, in total, the remaining global symmetry of the
vacuum is 
\begin{eqnarray}
 H = SU(N)\times U(N_f-N).
\end{eqnarray}
The moduli space of the vacua is
\begin{eqnarray}
{\cal M}_{\rm vac}= \frac{U(N_f)}{SU(N)\times U(N_f-N)}.
\label{vac1}
\end{eqnarray}

Next, let us gauge the $U(1)_B$ symmetry. 
If we gauge $U(1)_B$, the gauge group of the
magnetic theory becomes $U(N)$.\footnote{
Of course, this makes the original SQCD asymptotically non-free.
However, as a cut-off theory it could be 
useful for applications of the meta-stable vacua to
cosmologies and phenomenological model constructions.}
It is easy to see that the meta-stable vacua (\ref{VEV})
are still meta-stable vacua for the gauged $U(1)_B$ case.
The total global symmetry $G$
is in this case $G=SU(N_f)$. The remaining global symmetry 
$H$ is the same as before, but we can write it as 
$H=S[U(N_f-N)\times U(N)]= SU(N_f-N)\times SU(N)\times
\widetilde{U}(1)$, since $U(1)_B$ is gauged. 
Consequently, the vacuum manifold is given by 
\beq
\widetilde{\cal M}_{\rm vac} 
&=& \frac{SU(N_f)}{S[U(N)\times U(N_f-N)]} 
= \frac{SU(N_f)}{SU(N)\times SU(N_f-N)\times \widetilde{U}(1)} 
\nonumber\\
&=& Gr_{N_f,N}
\label{vac2}
\eeq
which is a Grassmanian manifold.

We can find the moduli space $\cal M_{\rm vac}$ easier 
by considering gauge invariant operators.
Here the gauge invariant operators are the meson $M_i^j$
and baryons $b_{i_1 i_2 \cdots i_N}$,  $\tilde{b}_{i_1 i_2 \cdots i_N}$ 
where $i_k$ runs from $1$ to $N_f$.
Actually, the global symmetry $U(N_f) \sim SU(N_f) \times U(1)_B$
was broken by the gauge invariant operators
$b_{1 2 \cdots N } \sim \mu \; \epsilon_{12 \cdots N}
\sim \tilde{b}_{1 2 \cdots N }$, 
the other components of the $b$ and $\tilde{b}$ vanish,  
and $M_i^j=0$.
Therefore, we find again that the unbroken global symmetry is 
$SU(N) \times U(N_f-N)$
and the moduli space ${\cal M}_{\rm vac} $ is indeed (\ref{vac1}),
because the moduli space is spanned by massless scalars
while we know that the massless scalars in the meta-stable vacua
are only the Nambu-Goldstone modes associated with the global symmetry
breaking \cite{ISS}. 
For the gauged $U(1)_B$ case,
the global symmetry becomes $SU(N_f)$ and 
the baryon $b$ is not a gauge invariant operator, but $ b \tilde{b}$ is.
Thus, the unbroken global symmetry group in this case 
is $S(U(N) \times U(N_f-N))$
and the moduli space $\widetilde{\cal M}_{\rm vac}  $ is indeed
given by (\ref{vac2}). 
In the limit of the vanishing $U(1)_B$ gauge coupling $e \rightarrow 0$, 
the scalar fields which are massive due to the super Higgs mechanism
become massless, which change 
$\widetilde{\cal M}_{\rm vac}$ to ${\cal M}_{\rm vac}$.
However, this is globally nontrivial because 
${\cal M}_{\rm vac} \neq \widetilde{\cal M}_{\rm vac}  
\times (S^1 \;{\rm of }\;U(1)_B)$.

Finally, we briefly mention
the moduli space of the meta-stable vacua in
the massive SQCD with the $SO(N_c)$ gauge group,
under a condition $0< N_c -4 < N_f < \frac{3}{2} (N_c-2)$.
The gauge group of the magnetic theory is $SO(N)$ ($N=N_f-N_c+4$), and 
the moduli space is 
\begin{eqnarray}
{\cal M}_{SO}= \frac{SO(N_f)}{SO(N) \times SO(N_f-N)}.
\label{somodu}
\end{eqnarray}
This is obtained in quite a similar way, since the structure of the
superpotential is the same. The $SO(N)$ appearing in the denominator is
from the gauge symmetry locked with a part of the global symmetry, thus 
the vacua are in the color-flavor locking phase.

\subsection{Homotopy groups}

Homotopy groups directly indicate the existence of global (or sigma
model) solitons, and we here evaluate the homotopy groups of the vacuum
manifold (\ref{vac1}), (\ref{vac2}) and (\ref{somodu}). 

First, we consider the case of 
the un-gauged $U(1)_B$ symmetry, (\ref{vac1}). 
The homotopy exact sequence \cite{exact} concerning the vacuum manifold 
${\cal M}_{\rm vac}=G/H$ is
\begin{eqnarray}
0= \pi_2(G)\stackrel{f_1}{\to} \pi_2(G/H) 
\stackrel{f_2}{\to} \pi_1(H) 
\stackrel{f_3}{\to} 
\pi_1(G) 
\stackrel{f_4}{\to} \pi_1(G/H) \stackrel{f_5}{\to} \pi_0(H)=0.
\end{eqnarray}
Here we have used the fact that $\pi_2(G)= \pi_2(U(N_f))=0$ and 
$\pi_0(H)=\pi_0(SU(N)\times U(N_f-N))=0$. We know that 
$\pi_1(U(M))={\bf Z}$ for any $M\geq 1$, so the exact sequence is 
written as 
\begin{eqnarray}
0 \stackrel{f_1}{\to} \pi_2(G/H) \stackrel{f_2}{\to} {\bf Z}
\stackrel{f_3}{\to}
{\bf Z} 
\stackrel{f_4}{\to} \pi_1(G/H) 
\stackrel{f_5}{\to}0.
\end{eqnarray}
To obtain the homotopy groups $\pi_i(G/H)$ for seeing the
solitons, it is necessary to know how the map ${\bf Z}\to {\bf Z}$ is 
organized. First, let us see how the $\pi_1(G)={\bf Z}$ is
generated. In the $G=U(N_f)$, consider the following loop $u(t)$
where $t$ is 
parameterizing the loop as $0\leq t \leq 1$ and $u(0)=u(1)$:
\begin{eqnarray}
& u(t) = e^{4\pi i t/N_f}{\bf 1}_{N_f} \in U(1)_B
\quad &(0\leq t \leq 1/2) \\
& u(t) \in SU(N_f) \quad &(1/2\leq t \leq 1)
\end{eqnarray}
This is a non-trivial loop going from the origin, through 
$e^{\pi i/N_f} {\bf 1}_{N_f}$, back to the origin. 
In the same manner, $\pi_1(H)={\bf Z}$ is generated
by a loop $u'(t)$ in the $U(N_f-N)$, whose definition is completely
analogous to $u(t)$. The map ${\bf Z}\to {\bf Z}$ is determined by the
relation between this $u'(t)$ and $u(t)$. Note that $u'(t)$ ($0\leq t
\leq 1/2$) is decomposed into a product of $U(1)_B$ and $SU(N_f)$ as  
\begin{eqnarray}
 u'(t) = 
\left(
\begin{array}{cc}
{\bf 1}_N & \\ & e^{4\pi i t/(N_f-N)}{\bf 1}_{N_f-N}
\end{array}
\right)
=
e^{4\pi t/N_f}
\left(
\begin{array}{cc}
e^{-4\pi i t/N}{\bf 1}_N & \\ 
& e^{4\pi i t/(N_f-N)}{\bf 1}_{N_f-N}
\end{array}
\right).
\nonumber
\end{eqnarray}
Looking at the $U(1)_B$ factor in the last expression, we find that the
loop $u'(t)$ reaches 
the point $e^{2\pi t/N_f}$ at $t=1/2$. This is the same point as
$u(1/2)$. Therefore, the loop $u(t)$ rounds once when the loop $u'(t)$
rounds once: the map ${\bf Z}\to {\bf Z}$ is one-to-one and on-to. 

Using this fact, we can obtain the homotopy groups of the quotient
space, using the exactness of the sequence.
First, using Ker($f_3$)$=0$, we have Im($f_2$)$=0$, which implies
Ker($f_2$)$=\pi_2(G/H)$. Again the exactness means 
Ker($f_2$)$=$Im($f_1$)$=0$, thus we obtain $\pi_2(G/H)=0$.
On the other hand, since Im($f_3$)$={\bf Z}$, we have 
Ker($f_4$)$={\bf Z}$. Hence Im($f_4$)$=0$, which leads to
Ker($f_5$)$=0$. However, $f_5$ is the last element 
in the exact sequence, so 
$\pi_1(G/H)=$Ker($f_5$). Thus we obtain $\pi_1(G/H)=0$. This is the
proof of the result (\ref{hom1}).

The moduli space of the theory with the gauged $U(1)_B$ symmetry 
has a non-trivial homotopy, 
\begin{eqnarray}
\pi_2
\left( \frac{SU(N_f)}{SU(N_f-N)\times SU(N) 
\times \widetilde U(1)} \right)
\qquad\qquad\nonumber\\
= \pi_1 \left( SU(N_f-N) \times SU(N) \times \widetilde U(1) \right) 
= {\bf Z}.
\end{eqnarray}
We have used a well-known homotopy formula which is accessible in this
case. 

Finally, for the gauge group $SO(N_c)$, we write the moduli space
(\ref{somodu}) as $G/H$ where $G=SO(N_f)/SO(N_f-N)$ is a Stiefel
manifold and $H=SO(N)$. To obtain the homotopy groups of $G/H$, we use
the homotopy groups of $G$: 
$\pi_{i<N_f-N}(G) = 0$, $\pi_{N_f-N}(G)={\bf Z}$ (for even $N_f-N$ or
$N=1$), and $\pi_{N_f-N}(G)={\bf Z}_2$ (for odd $N_f-N$ and $N>1$.
Then, the exact sequence of the following generic form 
\begin{eqnarray}
0= \pi_i (G) \to \pi_i(G/H) \to \pi_{i-1}(H) \to \pi_{i-1}(G)=0
\end{eqnarray}
leads to a generic formula for homotopy groups of $G/H$ (a similar
derivation for $O(N_f)$ group can be found in \cite{exact}). 
But when $i$ is
large or $N_f$ is small, the end points of the above exact sequence do
not vanish, and special treatment is required. The results are already
listed at the end of section \ref{sec:topsol}. In addition, for example,
one can find $\pi_3(G/H)=0$ for $N>3$ or $N_f-N>3$.

\section{Brane Realization of Vortex Strings in Meta-Stable Vacua}
\label{sec:branestring}
\setcounter{footnote}{1}

In this section, we show that classically there exist solitonic strings
in the magnetic side of the SQCD, by studying corresponding brane
configurations. 

In section \ref{sec:issbrane}, 
we first review the brane realization of the meta-stable vacua
\cite{OO2, Franco:2006ht,Bena:2006rg}.
Then, in section \ref{sec:rev}, we review the brane
realizations of the ${\cal N}=2$ supersymmetric gauge theories and their
solitonic 
strings. 
In fact, in section \ref{sec:braneISSstring}, 
we will find a useful analogy 
between the solitonic strings in the supersymmetry-breaking meta-stable 
vacua and the well-known BPS solitonic strings in supersymmetric vacua
of the ${\cal N}=2$ supersymmetric gauge theory.
There we construct brane
configurations corresponding to the solitonic strings in the magnetic
dual of the massive SQCD. Various properties of the solitonic strings are
predicted from string theory.

\subsection{Review: Brane realization of meta-stable vacua in SQCD}
\label{sec:issbrane}

\begin{figure}[ht]
\begin{center}
\begin{tabular}{ccc}
\includegraphics[width=7cm]{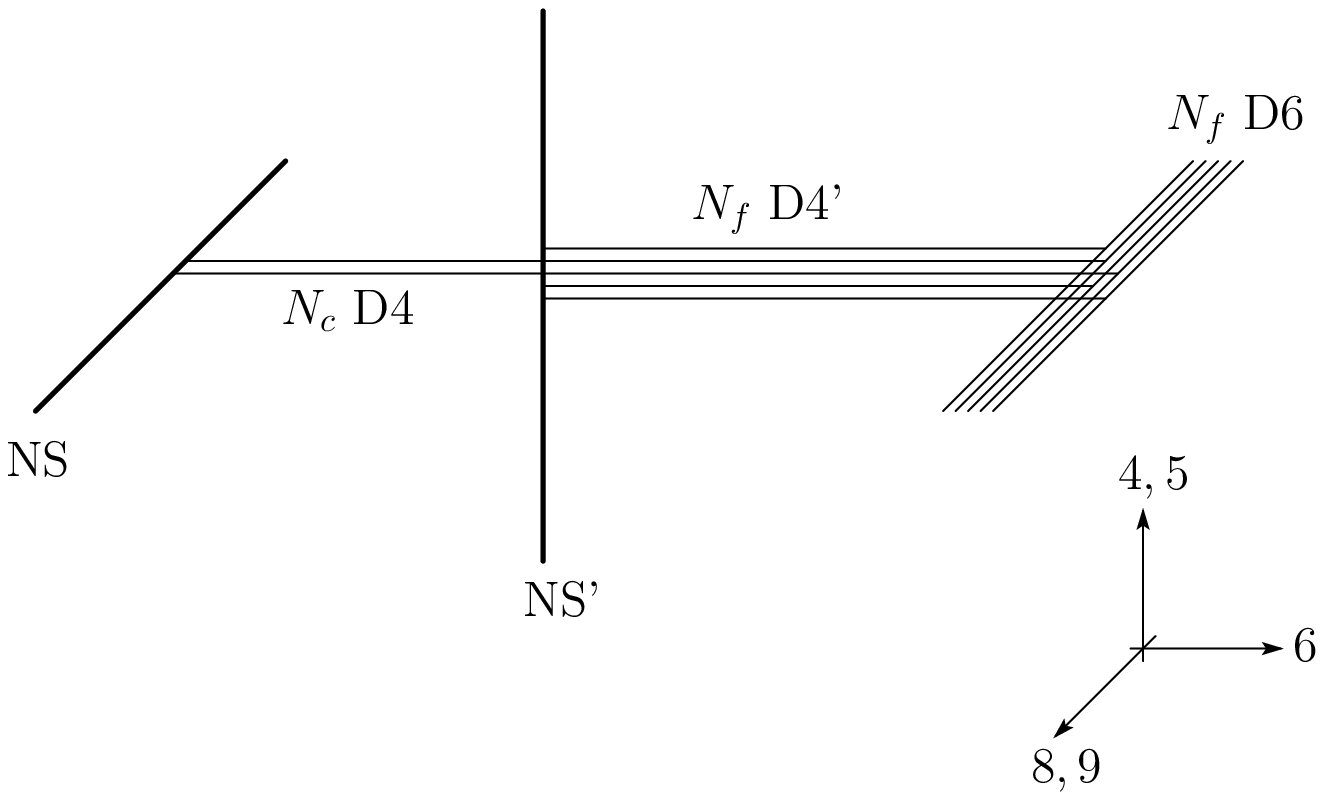}
&\ \ &
\includegraphics[width=6.5cm]{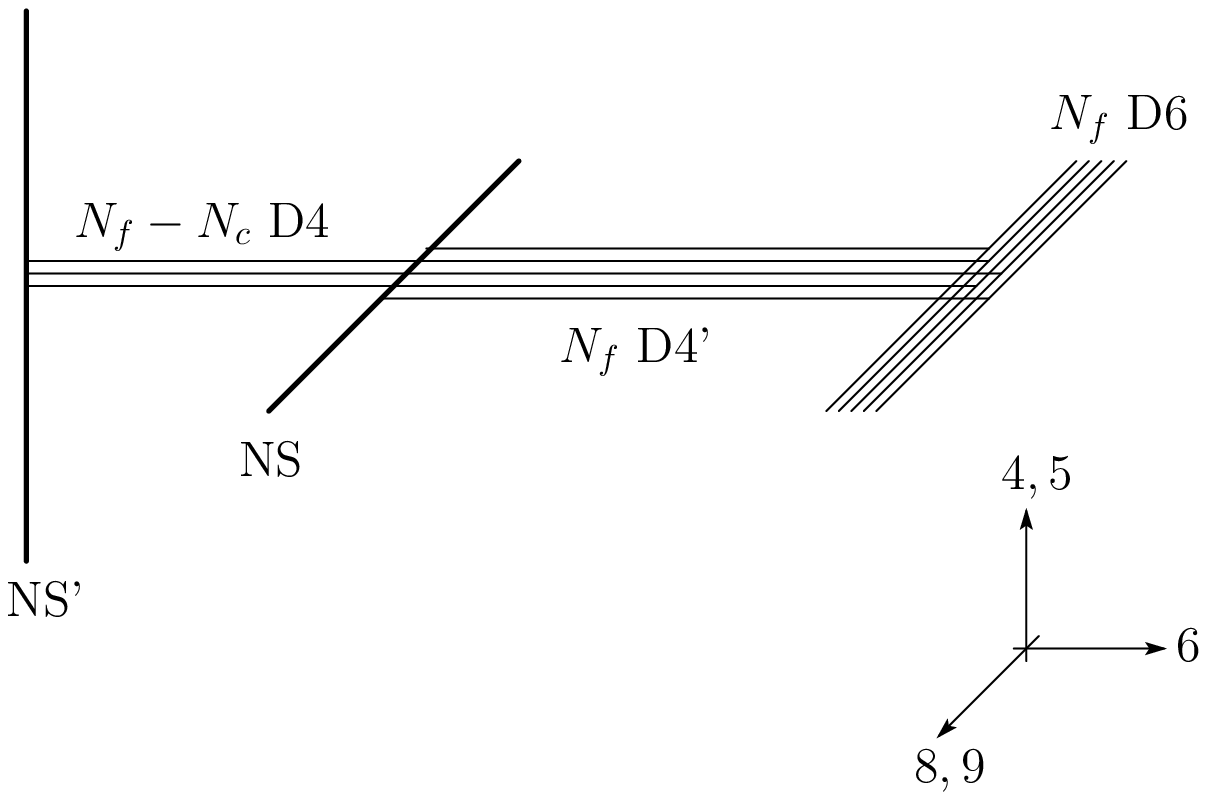}\\
(a) Electric theory & & (b) Magnetic theory
\end{tabular}
\caption{Brane configuration for massless SQCD}
\label{brane_SQCD}
\end{center}
\end{figure}

The brane configurations in the type IIA string theory can capture well
the properties of 
both the electric and the magnetic theories of the SQCD.\footnote{The
brane configurations are valid for analyzing the meta-stable vacua only in
the limit $g_s \rightarrow 0$ \cite{Witten:1997sc}.
In this paper, however, we will concentrate on classical solitons
in the meta-stable vacua, thus brane configurations are 
helpful.}
The Hanany-Witten setup \cite{Hanany:1996ie} for the ${\cal N}=1$ SQCD 
\cite{Elitzur:1997fh} (see \cite{Giveon:1998sr} for an extensive review)
consists of two NS5-branes,
$N_f$ D6-branes and $N_c$ D4-branes ($N_f$ D4-branes) whose
world-volume orientations are summarized in table \ref{wv:ht_n1}. 
The SQCD is realized on the D4-brane worldvolume at low energy.
\begin{table}[ht]
\begin{center}
\begin{tabular}{c|ccccccccc}
NS 	& 1 & 2 & 3 & -- & -- & -- & -- & 8 & 9 \\
NS' & 1 & 2 & 3 & 4 & 5 & -- & -- & -- & -- \\
D6 	& 1 & 2 & 3 & -- & -- & -- & 7 & 8 & 9 \\
D4 	& 1 & 2 & 3 & -- & -- & 6 & -- & -- & -- 
\end{tabular}
\caption{Hanany-Witten setup for the ${\cal N}=1$ supersymmetric gauge
 theory} 
\label{wv:ht_n1}
\end{center}
\end{table}
The brane configuration for the massless SQCD is
depicted in Fig.\ref{brane_SQCD}. The electric theory is realized 
in Fig.\ref{brane_SQCD}(a). The $U(N_c)$ vector multiplet corresponds to
the spectrum of a fundamental string ending on the $N_c$ D4-branes which
are suspended between the 
NS5-brane and the NS'5-brane,  while the chiral multiplets
$Q$ and $\widetilde Q$ come from a fundamental string stretched
between the $N_c$ D4-branes and the $N_f$ D4'-branes.
The dual (magnetic) theory can be obtained by exchanging the positions of
NS5-brane 
and NS'5-brane, for example, on the $x^6$ axis, see
Fig.\ref{brane_SQCD}(b). 
Hence, the $U(N)$ vector multiplet appears from a fundamental string
between the $N = N_f - N_c$ D4-branes, 
and the dual quarks $q$ and $\widetilde q$ come from fundamental strings
between the D4-branes and the D4'-branes. 
Furthermore, the meson field $M$ 
appears from fundamental strings on the $N_f$ D4'-branes,  which
corresponds to a 
massless degree of freedom for the transverse motion of the 
D4'-branes for the $x^8$ and the $x^9$ directions.

\begin{figure}[ht]
\begin{center}
\begin{tabular}{ccc}
\includegraphics[width=7.2cm]{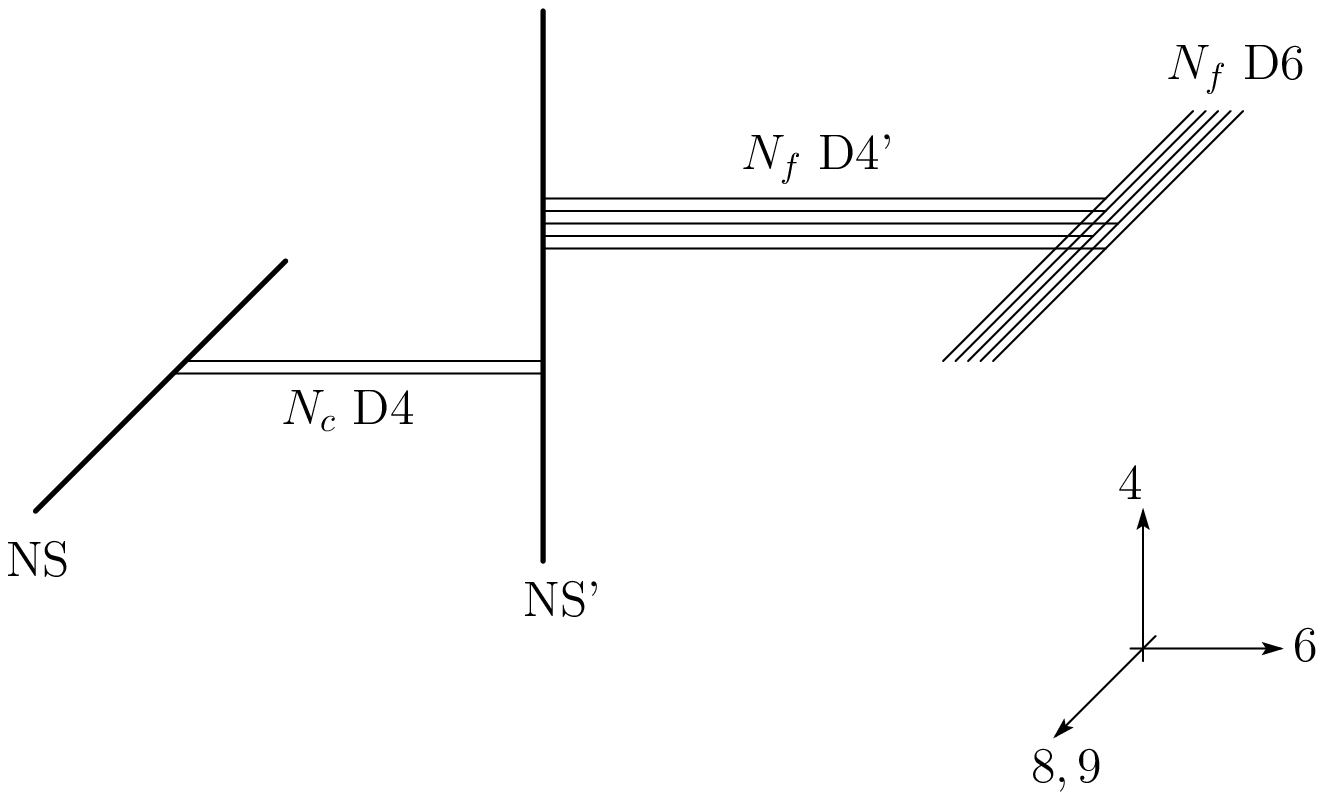}
&\ \ &
\includegraphics[width=6.7cm]{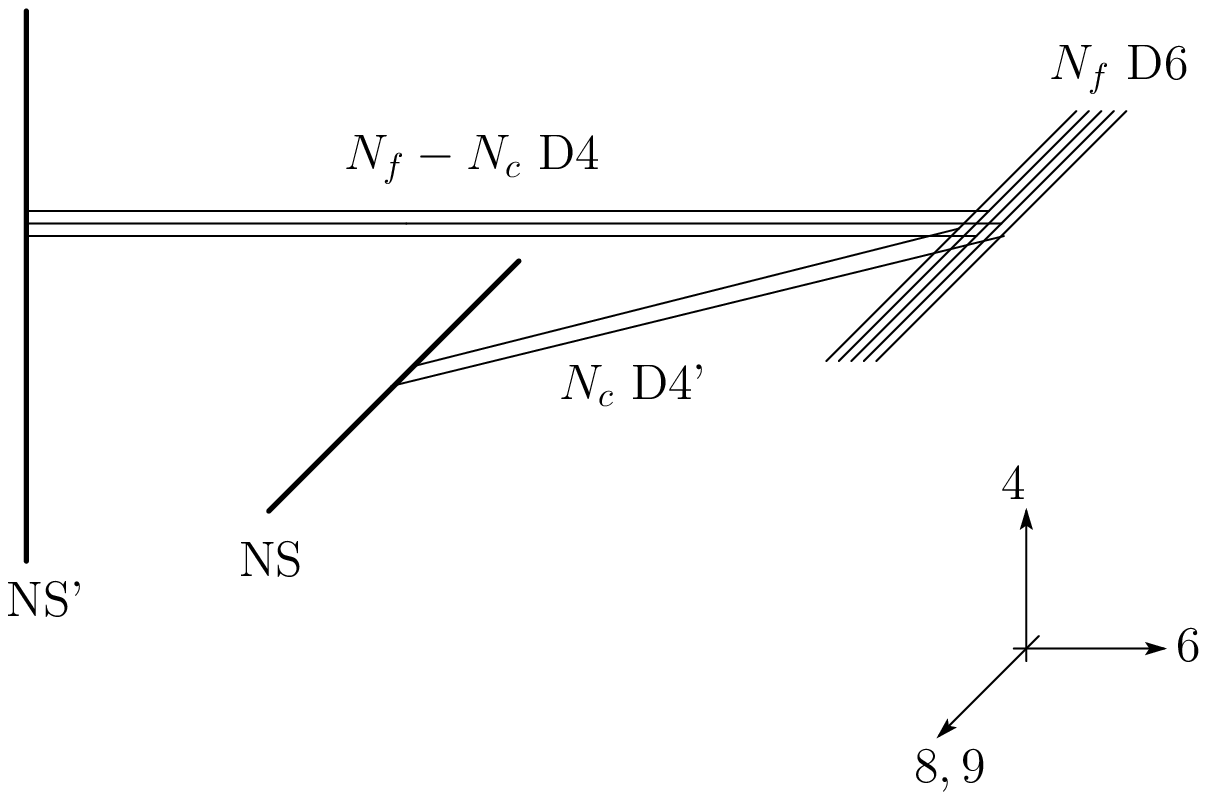}\\
(a) Electric theory & & (b) Magnetic theory
\end{tabular}
\caption{Brane configuration for massive SQCD}
\label{mn1}
\end{center}
\end{figure}

The brane configuration for the electric theory of the massive SQCD (the
masses are real) 
is obtained just by parallelly shifting the D4'-branes along the 
$x^4$ axis away from the D4-branes, as shown in Fig.\ref{mn1}(a).
Minimum length of a fundamental string stretched 
between the D4-branes and the D4'-branes is the 
distance between them. This non-zero distance
leads to nonzero masses for the quark fields $Q$ and $\widetilde Q$.
Let us next turn to the dual theory of the massive SQCD.
The brane configuration of it can provide an intuitive and good
understanding of the dual theory \cite{OO2,Franco:2006ht,Bena:2006rg}. 
We start
with Fig.\ref{brane_SQCD}(b) 
and we lift the $N_f$ D6-branes parallelly upward along the $x^4$ axis.
Then the $N (= N_f-N_c)$ D4-branes and the same number of D4'-branes 
in Fig.~\ref{brane_SQCD}(b) are
joined together 
and pulled by the D6-branes, then lifted upward. 
On the other hand, remaining $N_c (=N_f - N)$ D4'-branes cannot 
make a pair with any of the D4-branes, so they are still stretched
between the NS5-brane and the D6-branes, see Fig.\ref{mn1}(b)
\cite{OO2,Franco:2006ht,Bena:2006rg}. 
Obviously, the brane configuration in Fig.\ref{mn1}(b) breaks
the bulk supersymmetries completely,
and it is nothing but the supersymmetry-breaking meta-stable vacua.
Comparing Fig.\ref{brane_SQCD}(b) and Fig.\ref{mn1}(b), 
the length of the $N_f-N (= N_c)$ D4-branes in the former longer than
that in the latter. This difference is
regarded as the potential energy of the supersymmetry-breaking 
meta-stable vacua, and it agrees with (\ref{meta}).

As described in the last of section \ref{sec:iss}, we may introduce
various quark masses. In this case, the position of the D6-branes in the 
$x^4$-$x^5$ plane is specified by $m_i$. The situation of the
supersymmetry 
restoration at the vacuum, described in section \ref{sec:iss},
can be easily understood in the brane configuration. When $m_{N+1}=
\cdots = m_{N_f}=0$, both the NS5-brane and the $N_c$ D6-branes sit at
the origin of the $x^4$-$x^5$ plane, thus the $N_c$ D4'-branes connecting
them are aligned parallel to the remaining $N$ D4-branes. Thus the bulk
supersymmetries are not completely broken.

\subsection{Review: ${\cal N}=2$ supersymmetric gauge theory and vortex
  strings} 
\label{sec:rev}

In this subsection we review the brane configuration of BPS vortex
strings in ${\cal N}=2$ non-Abelian gauge theory \cite{Hanany:2004ea}
(see also \cite{Lee:1999ze} for brane configurations of 
vortices in Abelian gauge theory). 
This in fact
turns out to be quite helpful for constructing a brane configuration of
vortex strings in the supersymmetry-breaking meta-stable vacua in the
next subsection.

\vspace{10pt}
\noindent
{\it \underline{Brane realization and field theory vacua}}
\vspace{5pt}

\begin{figure}[t]
\begin{center}
\begin{tabular}{ccc}
\includegraphics[width=6cm]{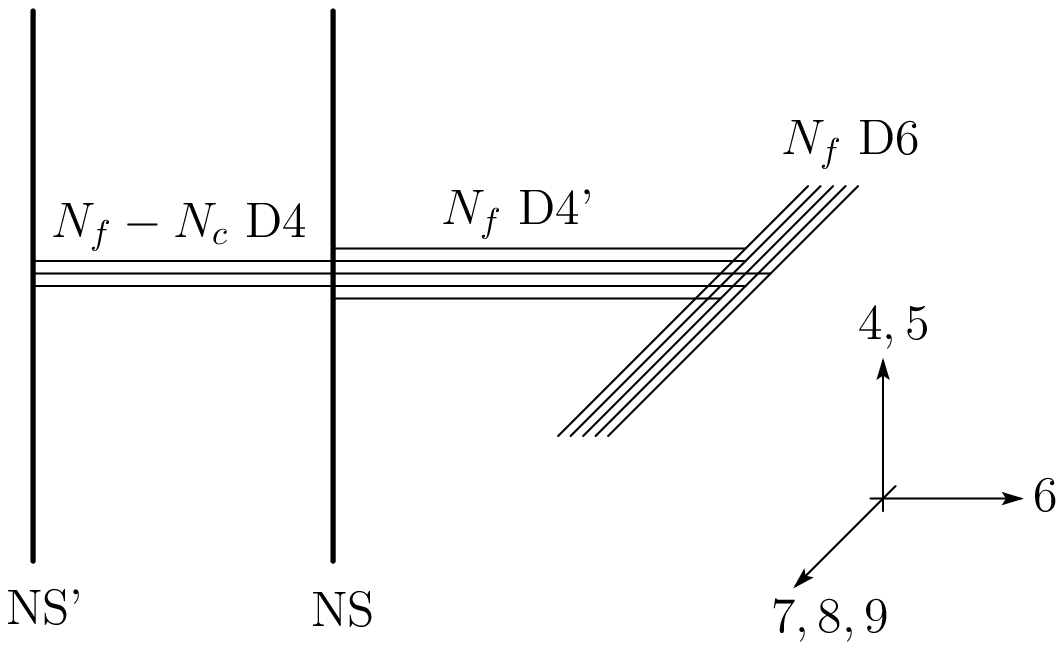}
&\qquad\quad &
\includegraphics[width=7.5cm]{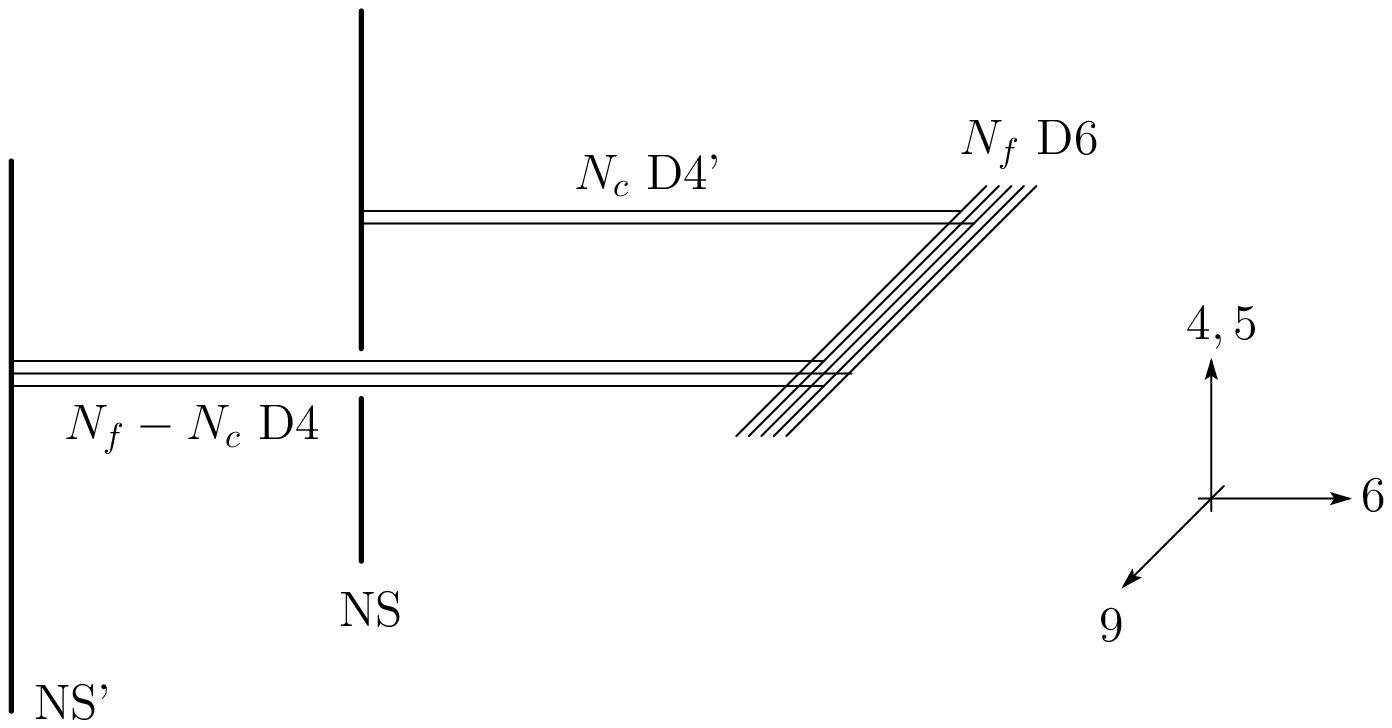}\\
(a) vanishing FI term & & (b) non-zero FI term
\end{tabular}
\caption{Brane realization of the 4d ${\cal N}=2$ supersymmetric $U(N)$
 gauge theories.} 
\label{ht_n2}
\end{center}
\end{figure}

We start with the ${\cal N}=2$ Hanany-Witten set up, whose brane
orientations are summarized in table \ref{wv:ht_n2}.
We consider the ${\cal N}=2$ $4d$ supersymmetric $U(N)$ gauge theory on
$N$ D4-branes suspended between a NS5-brane and a NS'5-brane, see
Fig.\ref{ht_n2}(a). 
\begin{table}[htb]
\begin{center}
\begin{minipage}{120mm}
\begin{center}
\begin{tabular}{c|ccccccccc}
NS 	& 1 & 2 & 3 & 4 & 5 & -- & -- & -- & -- \\
NS' & 1 & 2 & 3 & 4 & 5 & -- & -- & -- & -- \\
D6 	& 1 & 2 & 3 & -- & -- & -- & 7 & 8 & 9 \\
D4 	& 1 & 2 & 3 & -- & -- & 6 & -- & -- & -- \\
D2 	& -- & -- & 3 & -- & -- & -- & -- & -- & 9 
\end{tabular}
\caption{Hanany-Witten setup for the ${\cal N}=2$ $U(N)$ gauge
 theory. The D2-brane is added to describe the solitonic strings in the
 theory. }
\label{wv:ht_n2}
\end{center}
\end{minipage}
\end{center}
\end{table}
The fields which appear in the gauge theory are hypermultiplets
including scalar components of dual quarks $q$ and $\widetilde q$, and 
a vector multiplet including a complex scalar field $\Sigma$ in
the adjoint representation of the gauge group $U(N)$.
Here $q$ and $\widetilde{q}$ appear as in the same way as 
the magnetic theory of the SQCD. The new field  
$\Sigma$ comes from fundamental-strings whose end points are attached on
the  
$N$ D4-branes and it corresponds to transverse motion of the 
D4-branes along the $x^4$ and $x^5$ direction. 
On the other hand, the meson field $M$ in the magnetic theory of the 
SQCD does not appear here, because the $N_f$ D4'-branes cannot
move freely in this situation. 

We can turn on a Fayet-Iliopoulos (FI) 
parameter $v^2$ while keeping the ${\cal N}=2$
supersymmetries. This is necessary to have BPS solitonic strings.
In the brane configuration, the FI term is realized as a parallel
transport of the NS5-brane in the space of $x^7,x^8,x^9$. 
As an example, we parallel-transport 
the NS5-brane along the $x^9$ axis\footnote{The
rotation in the $x^7,x^8,x^9$ space corresponds in the field theory to
the $SU(2)_R$ symmetry. Namely,
the FI parameters are transformed as a triplet under the $SU(2)_R$.
In this paper, we assume that the parallel-transport 
of the NS5-brane along 
the $x^7$ axis corresponds to the FI $D$-term while
that along $x^8,x^9$ axis to for the FI $F$-term.},
see Fig.\ref{ht_n2}(b). 
The superpotential of this field theory is quite similar to
that given in equation (\ref{superpotential}) of the magnetic theory of
the SQCD:
\beq
W = g{\rm Tr}_c \left[ q \Sigma \widetilde q \right] - v^2 {\rm Tr}_c \Sigma.
\eeq

With this FI parameter $v^2$, as is obvious in Fig.\ref{ht_n2}(b), the
D4-branes cannot move freely. This fact can be seen consistently in the
field theory, by looking at the scalar potentials
\beq
V_F &=& {\rm Tr}_c
\left\{
g^2 \left|q \widetilde q - v^2 {\bf 1}_N \right|^2
+ |q \Sigma|^2 + |\widetilde q^\dagger \Sigma^\dagger|^2
\right\},\\
V_D &=& {\rm Tr}_c
\left\{
\frac{g^2}{4}\left(qq^\dagger - \widetilde q^\dagger \widetilde q \right)^2
- \frac{1}{g^2}\left[\Sigma,\Sigma^\dagger \right]^2
\right\}.
\eeq
These scalar potentials are almost the same as those in the SQCD
given in equations (\ref{iss_pot_1}) and (\ref{iss_pot_2}).
The only difference is the size of the field $\Sigma$ $(N\times N)$ and 
the field $M$ $(N_f\times N_f)$, and consequently the rank condition.
The classical vacuum of this ${\cal N}=2$ model is in the Higgs phase:
\beq
\Sigma = 0,\quad 
q q^\dagger - \widetilde q^\dagger \widetilde q = 0,
\quad
q \widetilde q = v^2 {\bf 1}_{N_c}.
\eeq
There is no flat direction in $\Sigma$ at the classical level 
on the contrary to the case of the ${\cal N}=1$ SQCD. 
Furthermore, the Higgs branch is well-known to be a
cotangent bundle over a complex Grassmanian manifold
\beq
T^\star Gr_{N_f,N} = T^\star \left[
\frac{SU(N_f)}{SU(N) \times SU(N_c) \times \widetilde{U}(1)}
\right].
\eeq
The base space Grassmanian is parameterized by $q = \widetilde q^\dagger$
similarly to the case of the ${\cal N}=1$ SQCD.
In this case the rank condition is satisfied because the rank
of $q\widetilde q$ is the same as $v^2{\bf 1}_N$, so that the vacuum
energy vanishes. Thus the vacua maintain the full supersymmetries.

\vspace{15pt}
\noindent
{\it \underline{1/2 BPS solitonic strings}}
\vspace{5pt}

There are 1/2 BPS solitonic strings (vortex strings) in this 
${\cal N}=2$ theory. They are called semilocal 
non-Abelian vortex strings. The 
Abelian version has been known for decades \cite{Vachaspati:1991dz},
while its non-Abelian extension of our concern has been considered 
in \cite{Hanany:2004ea} (see also \cite{Auzzi:2003fs,Auzzi:2003em,
Shifman:2004dr}).
The non-Abelian semilocal string is a natural extension of the
well-known
Abrikosov-Nielsen-Olesen (ANO) vortex in the Abelian-Higgs model. The
1/2 BPS equation for the vortex is 
\beq
(D_1 \pm iD_2)\phi = 0,\quad
F_{12} \pm \frac{g^2}{2}\left(2v^2 - \phi\phi^\dagger \right) = 0,
\label{bpseqano}
\eeq
where we have assumed that all the fields depend only on $x^1$ and $x^2$,
and 
\beq
q = \widetilde q^\dagger \equiv \frac{\phi}{\sqrt 2},\quad
\Sigma = A_0 = A_3 = 0,
\eeq
which are consistent with all the equations of motion.
The tension (the energy per unit length along $x^3$) of the system 
is bounded from below by the topological charge 
$k \in \pi_1(U(N))={\bf Z}$, 
and the bound is saturated by any solution of the 1/2 BPS equations
(\ref{bpseqano}),
\beq
{\cal E} = \mp 2v^2 \int d^2x\ F_{12} = 4v^2 \pi |k|.
\label{tension}
\eeq

As mentioned before, the vacuum shows the color-flavor locking,
namely the symmetry of the vacuum is $SU(N)_{c+f}$. When a minimal
vortex sits in the vacuum, the symmetry is spontaneously broken to
$S[U(N-1) \times U(1)]$, so the moduli space of the vortex includes
in particular an ``orientational'' moduli 
${\bf C}P^{N-1}$, in addition to the position moduli ${\bf C}$.
This ${\bf C} \times {\bf C}P^{N-1}$ is the total 
moduli space for the BPS non-Abelian strings with $N=N_f$, but
the 1/2 BPS semilocal vortex strings ($N_f > N$) have
additional moduli parameters concerning the size of the vortices.

\begin{figure}[ht]
\begin{center}
\includegraphics[width=9cm]{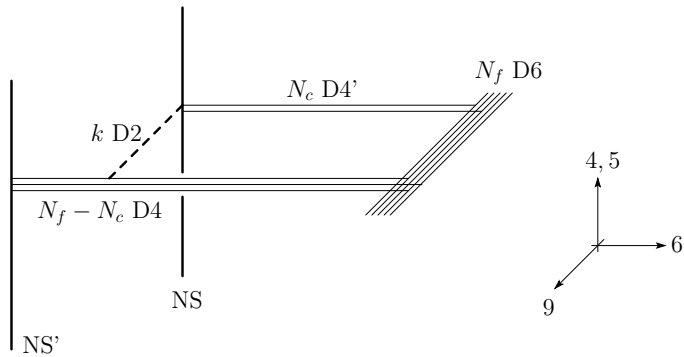}
\caption{1/2 BPS semilocal strings in ${\cal N}=2$ $4d$ super Yang-Mills-Higgs system}
\label{ht}
\end{center}
\end{figure}

This non-Abelian semilocal string can be realized in the brane 
configuration, which was found by Hanany and Tong
\cite{Hanany:2004ea}. 
The soliton with the co-dimension 2 (compared to the D4-branes) is $k$
D2-branes which are suspended between the 
NS'5-brane and the $(N_f-N_c)$ D4-branes, as shown in Fig.\ref{ht}.
The worldvolume of the D2-branes is along $x^0, x^1$ and $x^2$, as in
the table \ref{wv:ht_n2}.  
The tension of the D2-branes is proportional to the distance between
the NS5-brane and the NS'5-brane (along the $x^9$ axis), namely the
amount of the FI term $2v^2$. 
This is consistent with the field theory result in equation
(\ref{tension}). 

A topological property of the moduli space for 
these 1/2 BPS non-Abelian semilocal strings
is captured from massless excitations of fundamental
strings on  
$k$ D2-branes, $N$ D4-branes and $N_c$ D4'-branes \cite{Hanany:2004ea}.
Let us denote $k$ by $k$ matrix $Z$ for expected zero modes between D2-D2,
$k$ by $N$ matrix $\psi$ for those between D2-D4 and 
$N_c$ by $k$ matrix $\tilde \psi$ for those between D2-D4'.
Then the moduli space of the solitonic strings is given by
the following K\"ahler quotient:
\begin{eqnarray}
\left\{
\left[ Z, Z^\dagger \right] + \psi \psi^\dagger - \tilde \psi^\dagger \tilde \psi
\propto {\bf 1}_N
\right\}/U(k).
\end{eqnarray}
For a single vortex, $Z$ is just a complex constant. Then, if we
look at $\tilde \psi = 0$ sector, this quotient gives us the
${\bf C} \times {\bf C}P^{N-1}$ 
which is consistent with the above field theory result.
The 1/2 BPS non-Abelian semilocal strings have three kinds of moduli
parameters: 
(i) positions (ii) orientations (iii) sizes. Roughly speaking,
$Z$ corresponds to the positions, $\psi$ to the orientations and
$\tilde \psi$ to the sizes.

\subsection{Brane realization of vortex strings in meta-stable vacua}
\label{sec:braneISSstring}

We have seen that 1/2 BPS semilocal strings naturally appear in the
${\cal N}=2$ supersymmetric gauge theory with 
the FI parameter, and their corresponding D-brane picture is well
understood. In this subsection, we apply the idea to the magnetic theory
of the SQCD. We deal with two cases: magnetic theory of (i) 
massless ${\cal N}=1$ SQCD with an analogous FI term, 
and of (ii) massive ${\cal N}=1$ SQCD. 
We find D-branes
corresponding to the 
semilocal strings, which shows the existence of the solitonic strings in
the theories.

\vspace{20pt}
\noindent
{\it \underline{1/2 BPS solitonic strings in magnetic theory of massless
SQCD}} 
\vspace{5pt}

The brane configurations are quite useful to find out possible solitonic
defects, as reviewed in the previous subsection. Let us study what is a
possible introduction of the D2-brane in the brane configuration of the 
magnetic theory of the massless ${\cal N}=1$ SQCD, Fig.\ref{brane_SQCD}. 
It is clear that we cannot attach a D2-brane in this brane
configuration. But as is suggested from the ${\cal N}=2$ example in the
previous subsection, if we introduce a FI term, 
we obtain a stable vortex string, as we shall see below.

In the previous ${\cal N}=2$ case, we have three candidates, 
$x^7,x^8$ and $x^9$, as a possible direction along which we can 
parallel-transport
the NS'5-brane. In the present case of the magnetic theory of the
massless ${\cal N}=1$ SQCD, we have only the $x^7$ direction (FI $D$-term)
for the transport to maintain the supersymmetries of the vacua. 
The worldvolumes for the branes are summarized in table
\ref{table:string_n1}. 
\begin{table}[ht]
\begin{center}
\begin{minipage}{120mm}
\begin{center}
\begin{tabular}{c|ccccccccc}
NS 	& 1 & 2 & 3 & -- & -- & -- & -- & 8 & 9 \\
NS'     & 1 & 2 & 3 & 4 & 5 & -- & -- & -- & -- \\
D6 	& 1 & 2 & 3 & -- & -- & -- & 7 & 8 & 9 \\
D4 	& 1 & 2 & 3 & -- & -- & 6 & -- & -- & -- \\
D2 	& -- & -- & 3 & -- & -- & -- & 7 & -- & -- 
\end{tabular}
\caption{The Worldvolumes of the branes for the massless ${\cal N}=1$ 
SQCD and its dual. We added the D2-brane which represents a vortex
 string.} 
\label{table:string_n1}
\end{center}
\end{minipage}
\end{center}
\end{table}

\begin{figure}[ht]
\begin{center}
\begin{tabular}{ccc}
\includegraphics[width=7cm]{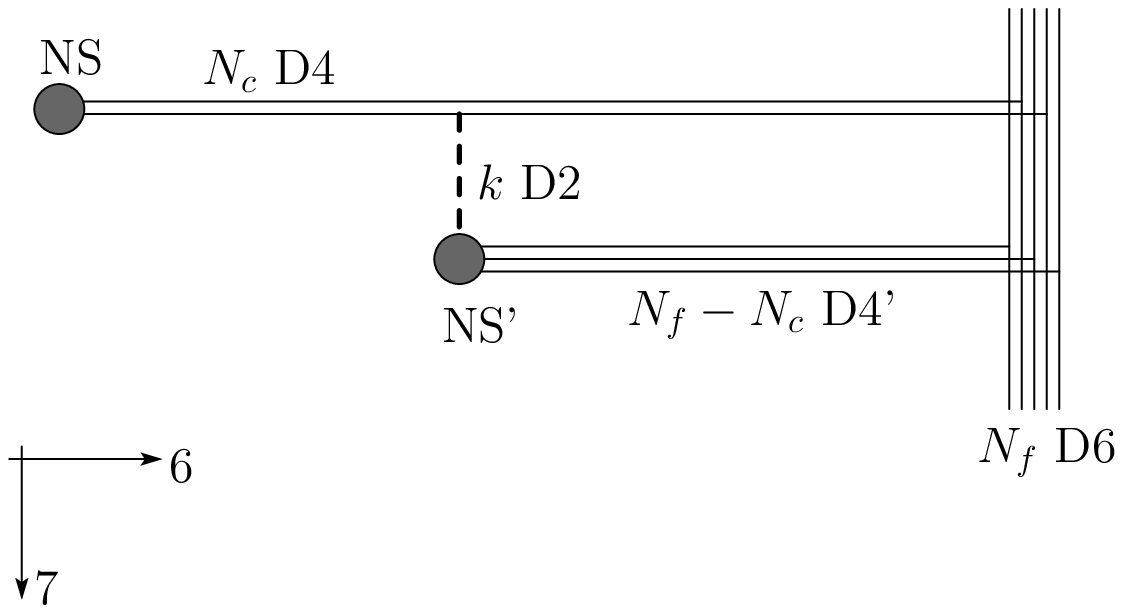}
& \qquad &
\includegraphics[width=7cm]{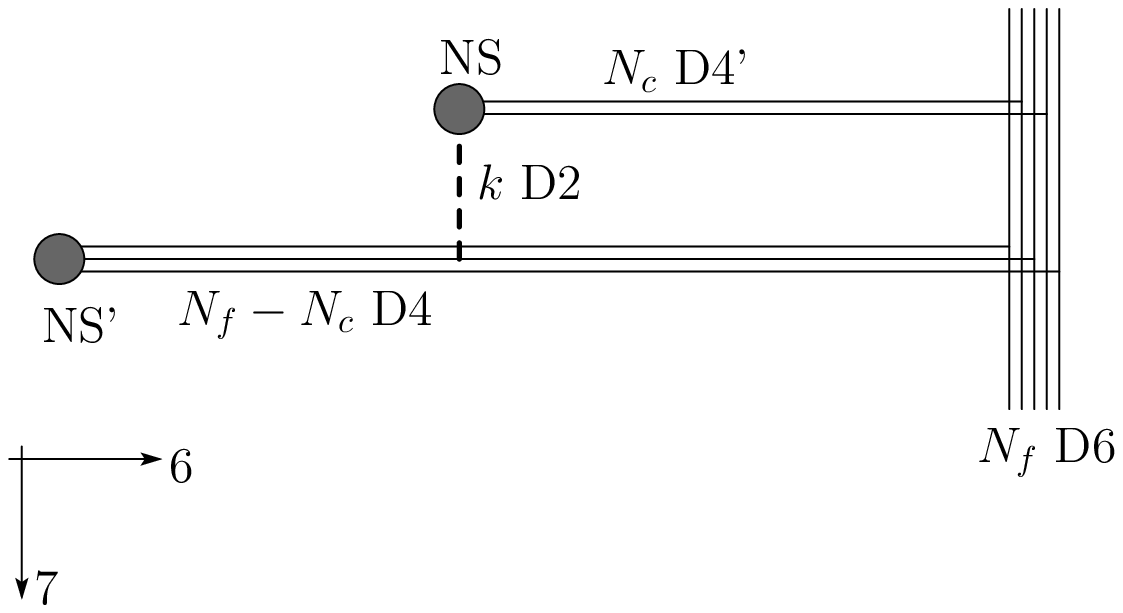}\\
(a) Electric side && (b) Magnetic side
\end{tabular}
\caption{Brane configurations for the 1/2 BPS semilocal strings in
 ${\cal N}=1$ SQCD.} 
\label{ht_n1}
\end{center}
\end{figure}

After turning on the FI $D$-term parameter (parallel-transporting the NS'5-brane
in the $x^7$ direction), 
$k$ D2-branes can stretch between the NS'5-brane and the 
$N_c$ D4-branes, see Fig.\ref{ht_n1}(a). This is a description 
of the electric side of the SQCD. In the magnetic side, 
we put the D2-branes stretched between the NS5-brane and 
the $(N_f - N_c)$ D4-branes, see Fig.\ref{ht_n1}(b).\footnote{
Field theoretical properties of the semilocal strings and relations to
the brane configurations and to the Seiberg dualities are
studied in \cite{EKMNOVY}.}

\vspace{20pt}
\noindent
{\it \underline{Non-BPS solitonic strings in magnetic theory of massive
SQCD}} 
\vspace{5pt}

Let us consider the possibility of the existence of the 
solitonic strings in the magnetic theory of the 
{\it massive} ${\cal N}=1$ SQCD without the FI parameter,
explained in section \ref{sec:iss}.
The 1/2 BPS semilocal strings in the ${\cal N}=2$ model required
a non-vanishing FI term for them to exist. 
Instead,
as has been described, the SQCD has a non-zero ``quark mass'' $\mu^2$,
which behaves in the magnetic side quite similarly to the FI term.
This ``mass'' term in fact supports the solitonic string in the dual SQCD,
as we will see in the following.

\begin{figure}[ht]
\begin{center}
\begin{tabular}{cl}
\includegraphics[width=7cm]{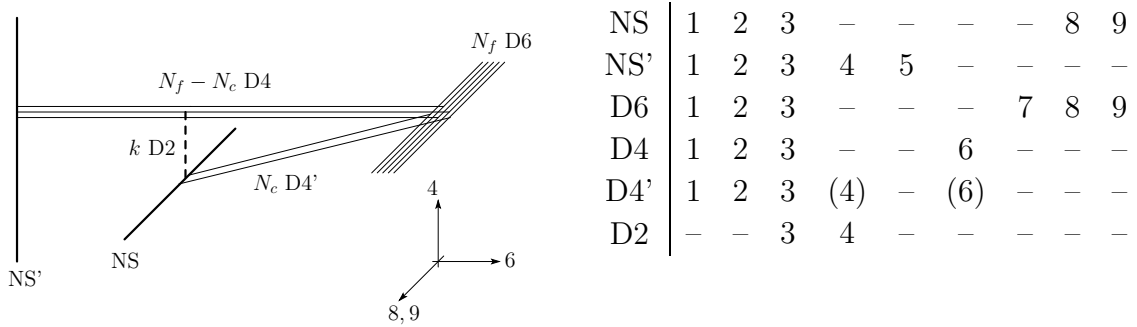}
&
\begin{minipage}{80mm}
\vspace*{-5cm}
\begin{center}
\begin{tabular}{c|ccccccccc}
NS 	& 1 & 2 & 3 & -- & -- & -- & -- & 8 & 9 \\
NS' & 1 & 2 & 3 & 4 & 5 & -- & -- & -- & -- \\
D6 	& 1 & 2 & 3 & -- & -- & -- & 7 & 8 & 9 \\
D4 	& 1 & 2 & 3 & -- & -- & 6 & -- & -- & -- \\
D4' 	& 1 & 2 & 3 & (4) & -- & (6) & -- & -- & -- \\
D2 	& -- & -- & 3 & 4 & -- & -- & -- & -- & -- 
\end{tabular}
\end{center}
\end{minipage}
\end{tabular}
\caption{Brane realization (D2-branes) of the solitonic strings in the
 meta-stable  vacua. }
\label{dht_n1}
\end{center}
\end{figure}

We can easily find a brane configuration for the solitonic string, 
by applying the idea of \cite{Hanany:2004ea}. 
In the brane realization of the 
magnetic theory provided by \cite{OO2,Franco:2006ht,Bena:2006rg}, 
we put a D2-brane suspended between the NS5-brane and the 
$(N_f-N_c)$ D4-branes. See Fig.\ref{dht_n1}. 
Because the D2-brane tends to minimize
its length, it is perpendicular to both the NS5-brane and the D4-branes.

From the brane configuration, we can extract several properties of the
solitonic strings. All of those are found to be consistent with 
our field theory analyses which will be presented in the next section. 
\begin{itemize}

\item {\bf Tension of the string.} The length of the D2-brane measures
      the tension. It is found to be proportional to $\mu^2$, the quark
mass in the original SQCD, 
since the length is identical to the distance between the D6-branes
and the NS5-brane in the $x^4$ axis. 

\item {\bf Supersymmetries. }
The meta-stable vacua completely break supersymmetries by themselves.
This has been realized as the tilted $N_c$ D4'-branes in the brane
configuration \cite{OO2,Franco:2006ht,Bena:2006rg}. 
We have to stress that the semilocal strings should also break the
supersymmetries completely, 
since the $N_f$ D6-branes along the directions $0123789$ and the 
D2-brane along $034$ are incompatible with supersymmetries.
So the solitonic string in the supersymmetry-breaking meta-stable vacua
      are 
non-BPS ($\equiv$ supersymmetry-breaking)\footnote{
Note that in our terminology, ``non-BPS'' means supersymmetry-breaking,
      and not the saturation of the Bogomol'nyi bound.}
semilocal strings. 

\item {\bf Existence of non-BPS strings in supersymmetry-preserving
      vacua.} 
Before proceeding to the field theory analysis, we study brane
configurations of non-BPS solitonic strings, not 
in the supersymmetry-breaking
vacua, but in the supersymmetry-preserving vacua, in the magnetic theory
of the ${\cal N}=1$ massive SQCD. 
\begin{figure}[t]
\begin{center}
\begin{tabular}{ccc}
\includegraphics[width=7cm]{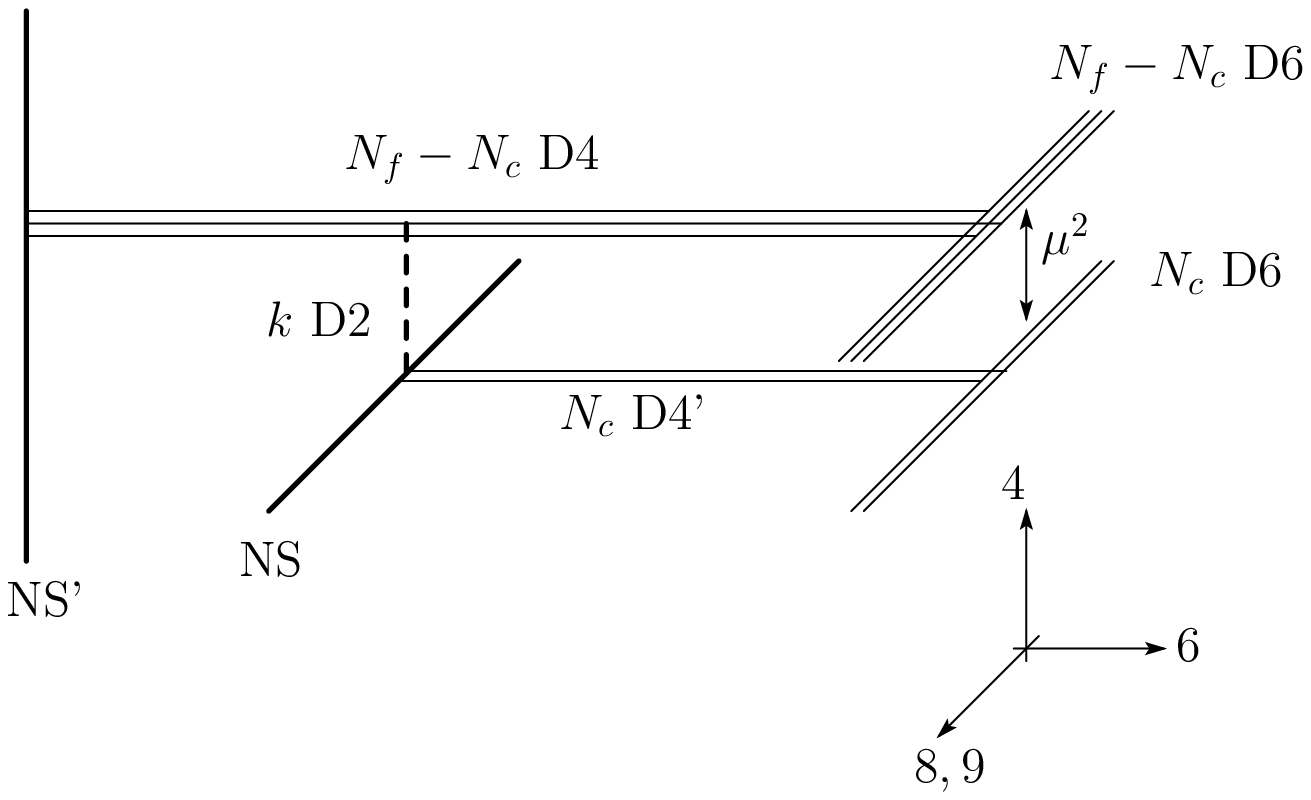} & \quad &
\includegraphics[width=7cm]{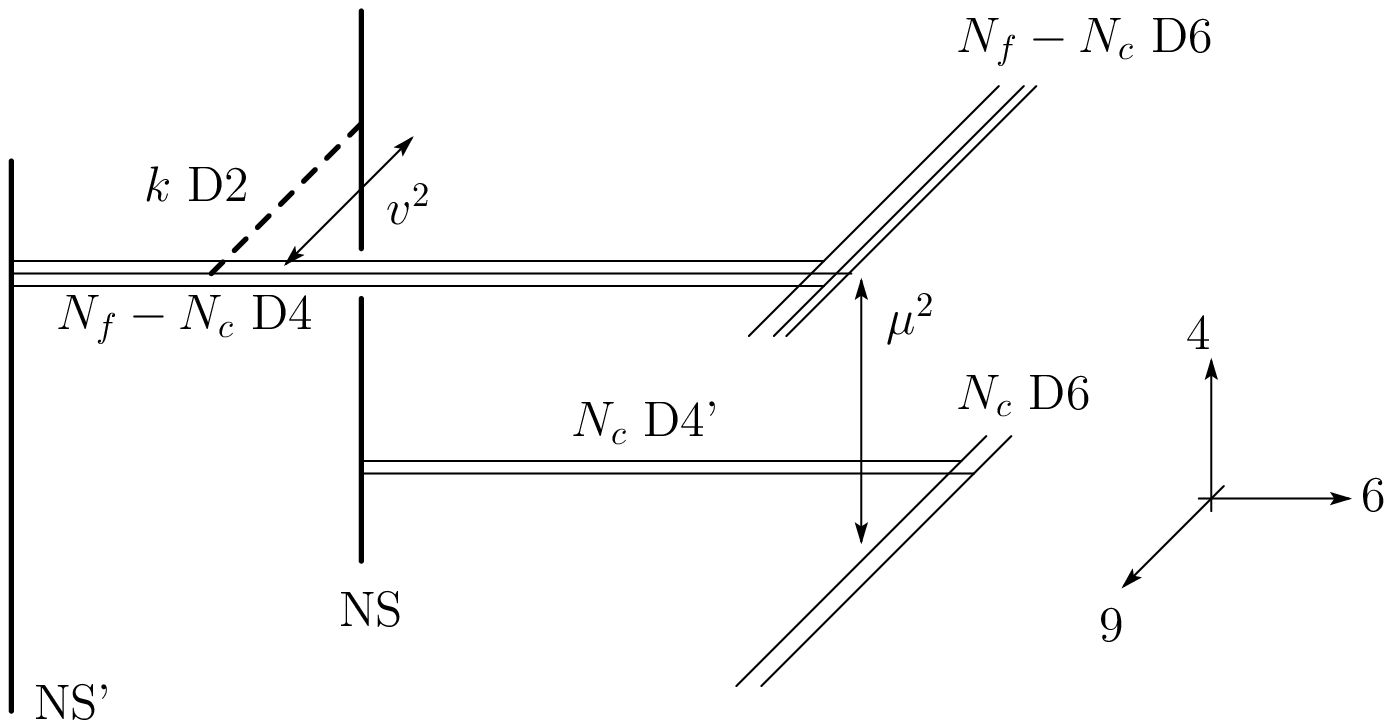}\\
(a) ${\cal N}=1$& & (b) ${\cal N}=2$
\end{tabular}
\caption{Non-BPS solitonic strings in massive dual ${\cal N}=1$ 
SQCD}
\label{smdht_susy_massive}
\end{center}
\end{figure}
As reviewed briefly at the end of section \ref{sec:iss}, 
if we replace the ``mass'' term $\mu^2 {\rm Tr}M$ in the superpotential
(\ref{superpotential}) by ${\rm Tr} [mM]$ where 
$m={\rm diag} (\mu^2,\cdots,\mu^2, 0, \cdots, 0)$,
the resultant vacuum becomes supersymmetric. 
The solitonic strings are again realized D2-branes between $N_f-N_c$
      D4-branes 
and NS5-brane. 
It appears that the brane configuration preserves some supersymmetries, 
but note that the D2-branes and the D6-branes are still 
in conflict with any compatible supersymmetries.
Thus the solitonic string in this supersymmetric vacuum is again
      non-BPS. 

The solitonic string in this vacuum is not a semilocal string but 
a non-Abelian string with the orientational moduli ${\bf C}P^{N-1}$,
because there is no global symmetry $SU(N_f)$ from the first place.
The brane configuration in Fig.\ref{smdht_susy_massive}(a) is 
quite similar to that in Fig.\ref{ht}. However, properties of 
the solitonic strings,
namely D2-branes suspended D4-branes and NS5-brane, are not so similar:
The solitonic strings in Fig.\ref{ht} are 1/2 BPS semilocal
      strings 
which have orientational moduli and size moduli, while those in 
Fig.\ref{smdht_susy_massive}(a) are non-BPS non-Abelian strings
which doesn't have the size moduli.

To illustrate this distinction, we draw an ${\cal N}=2$ D-brane
      configuration Fig.\ref{smdht_susy_massive}(b) which is much more
      similar  
to the brane configuration Fig.\ref{smdht_susy_massive}(a). 
This Fig.\ref{smdht_susy_massive}(b) shows a massive Hanany-Tong setup 
in which
the vacuum is not degenerate, 
since the non-degenerate masses break $SU(N_f)$ down to $SU(N)$
explicitly. So
the solitonic string in Fig.\ref{smdht_susy_massive}(b) is not a 1/2 BPS
semilocal non-Abelian string but a 1/2 BPS non-Abelian string without
the size moduli.

\item {\bf Existence of multi-tension strings.}

In general, we can put non-degenerate ``masses'' 
$m = {\rm diag}(m_1,m_2,\cdots,m_{N_f})$.
The corresponding D-brane configuration is shown in Fig.\ref{gene}.
Stable vacua are those in which $N_f - N_c$ horizontal D4-branes connect
the NS'5-brane and the $N_f - N_c$ D6-branes associated with 
the $N_f - N_c$ large masses $\{m_1,\cdots,m_{N_f-N_c}\}$.
The remaining $N_c$ D4'-branes are suspended between the NS5-brane
and the $N_c$ D6-branes associated with the light masses 
$\{m_{N_f-N_c+1},\cdots,m_{N_f-1},m_{N_f}\}$.
\begin{figure}[t]
\begin{center}
\includegraphics[width=9cm]{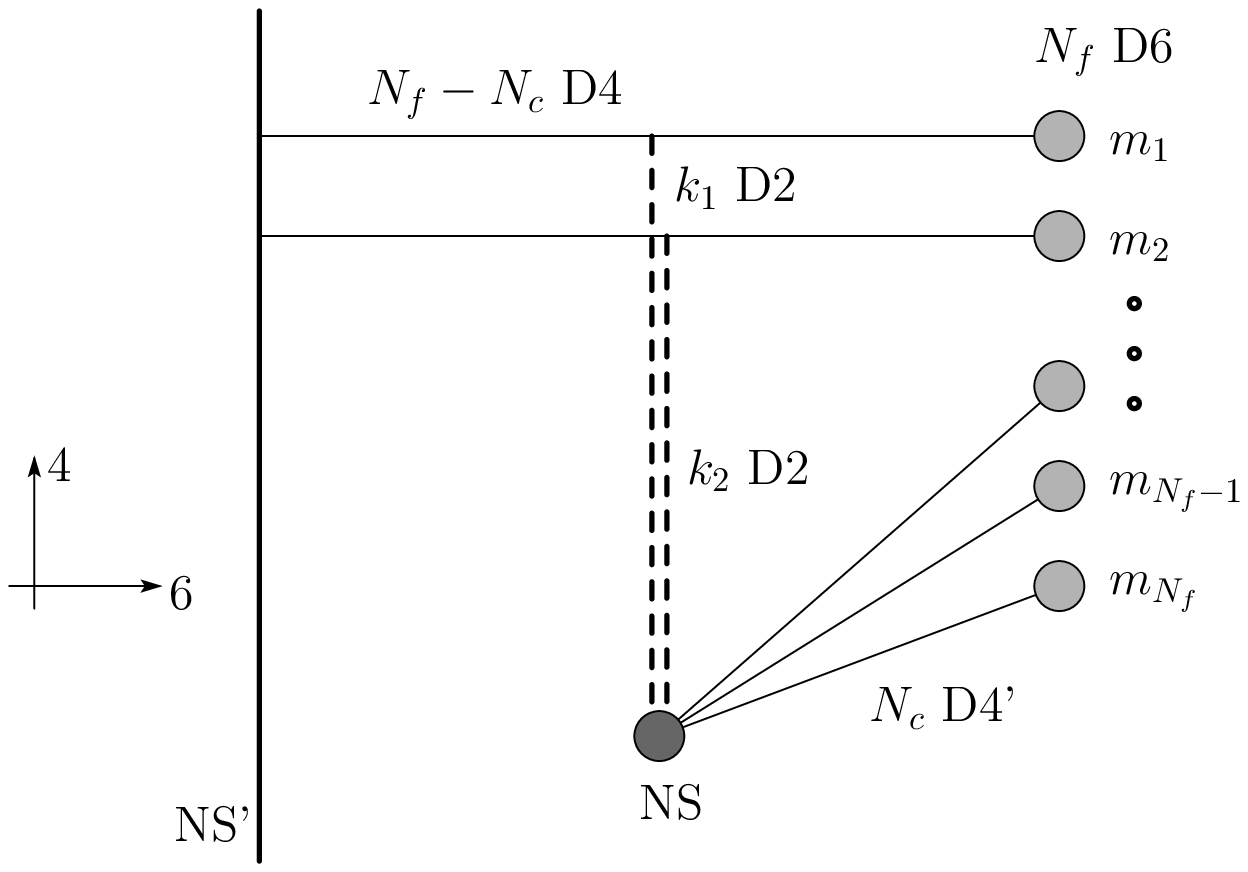}
\caption{Brane configuration of various strings with different tensions.}
\label{gene}
\end{center}
\end{figure}
Obviously one can put the D2-branes between the horizontal D4-branes 
and the NS5-brane. There are $N_f - N_c$ kinds of D2-branes,
depending on which D4-brane (labeled by $i=1,\cdots,N_f-N_c$)
the D2-brane ends on.
The tension of the solitonic string comes from the distance
between the $i$-th D4-brane and the
NS5-brane, hence is proportional to $m_i$. 
\end{itemize}

In the next section, we will give explicit solutions of the solitonic
strings in the magnetic theory of the SQCD, and show that indeed these
properties are equipped with the strings.

\section{Vortex String Solutions in Meta-Stable Vacua}
\label{sec:solution}
\setcounter{footnote}{1}

In this section, we explicitly construct the solitonic string
solutions of the equations of motion in the magnetic theory of the
massive SQCD. We find that the resultant strings have the properties
expected from the brane configurations: the tensions, the
supersymmetries, and the various species of the strings.

\subsection{Classical solutions}
\label{sec:classical sol}

The existence of the solution to the equations of motion can be seen
by considering the following field configurations
\beq
M = 0,\quad
 \frac{1}{\sqrt{2}}(q + \widetilde{q}^\dagger) \equiv \phi,
\quad 
\frac{1}{\sqrt{2}}(q - \widetilde{q}^\dagger) \equiv  \widetilde{\phi}
=0.
\label{assumption}
\eeq
Interestingly, this assumption leads to a system which is almost
identical to that of the semilocal strings. The only difference is the
vacuum energy, as we will see shortly below: 
in our case the vacuum energy is non-vanishing and thus
all the supersymmetries are always broken, while in the usual case of the
BPS semilocal strings some supersymmetries are preserved at the vacuum. 

With the truncation (\ref{assumption}),  
the Lagrangian takes the form\footnote{
When we take $g^2 \to \infty$ and simultaneously
$|h| \to \infty$, this model becomes a non-linear sigma model whose
target space is
\beq
{\cal M}_{\rm target} = 
\left\{ \phi^\dagger \phi - 2\mu^2 {\bf 1}_N \right\}/
\left[SU(N) \times U(1)_B\right] \simeq Gr_{N_f,N}.
\eeq
Note that in this limit we discard the infinite cosmological constant 
$(N_f - N)|h^2\mu^4| \to \infty$.
}
\beq
\widetilde {\cal L} = {\rm Tr}_c
\left[ - \frac{1}{2g^2} F_{\mu\nu} F^{\mu\nu}
- {\cal D}_\mu \phi {\cal D}^\mu \phi^\dagger
\right]
- \frac{|h^2|}{4}
{\rm Tr}_f \left[
\left(\phi^\dagger \phi - 2\mu^2 {\bf 1}_{N_f} \right)^2
\right],
\eeq
where $\phi$ is a complex matrix-valued field whose size is $N$ by
$N_f$. 
Here we gauged the $U(1)_B$ symmetry 
and unify it with the $SU(N)$ so
that the full gauge symmetry becomes $U(N)$, 
{\it i.e.} we put the gauge coupling of the $U(1)_B$ 
to be equal to that of the $SU(N)$.
Note that the $U(1)_B$ which we gauged corresponds to the original 
gauged baryonic $U(1)$ symmetry 
in the electric side and we took Seiberg's duality only for $SU(N_c)$.
Unifying $U(1)_B$ gauge coupling and $SU(N)$ gauge coupling here 
is just for a convinience. 
We will consider the generic case with non-coincident gauge couplings
in the Appendix.
Note that the above potential term can be written equivalently as 
\beq
\widetilde V=  |h^2\mu^4| (N_f-N)  
+ \frac{|h^2|}{4}
{\rm Tr}_c \left[
\left(\phi\phi^\dagger - 2\mu^2 {\bf 1}_{N} \right)^2
\right].
\label{eqsemi}
\eeq
As we will show shortly, the system is identical to that of the non-Abelian semilocal
strings plus the additive cosmological constant, the first term in
(\ref{eqsemi}). Because of this cosmological constant, all 
the supersymmetries
are always broken, although the system is similar to that of the
semilocal strings.
The equation of motion of the truncated model is 
\begin{eqnarray}
{\cal D}_\mu {\cal D}^\mu \phi = \frac{\partial \widetilde V}{\partial \phi^\dagger},
\quad
\frac{1}{g^2} {\cal D}^\mu F_{\mu\nu} =  -\frac{i}{2}\left(\phi {\cal D}_\nu \phi^\dagger
- {\cal D}_\nu \phi \phi^\dagger \right),
\label{2ndeom}
\end{eqnarray}
where $\mu,\nu = 1,2$ for the vortex strings extending along the $x^3$
axis. 
Notice that the cosmological constant in (\ref{eqsemi}) does not appear
in the equation of motion.
The equation of motion for $\widetilde{\phi}$ is satisfied with 
$\widetilde{\phi}=0$.

This equation of motion (\ref{2ndeom}) reduces to that for the 
well-known Abelian semilocal strings
when we choose $N_f>N=1$, and
their solutions were studied in detail \cite{Vachaspati:1991dz}.
We can find a minimal string solution of 
the non-Abelian semilocal string ($N_f>N\ge 2$)
for the equation (\ref{2ndeom})
by embedding the U(1) semilocal string solution in the model of
$N_f-N + 1$ flavors as
\begin{eqnarray}
F_{12} = 
\left(
\begin{array}{cccc}
0 & & & \\
& \ddots & & \\
& & 0 & \\
& & & F_{12}^\star
\end{array}
\right),\qquad
\phi = 
\left(
\begin{array}{ccccccc}
\sqrt 2 \mu & & & & & & \\
& \ddots & & & & & \\
& & \sqrt 2 \mu & & \\
& & & \phi^\star_0 
&\; \phi^\star_1\; & \cdots & \phi^\star_{N_f-N}
\end{array}
\right),
\label{embedding}
\end{eqnarray}
where we denoted $F_{12}^\star$ and 
$\{\phi^\star_a \;(a=0,1,\cdots,N_f\!-\!N)\}$ 
as solutions
of the minimal winding U(1) semilocal string.
This is the solution of the solitonic strings in the
supersymmetry-breaking meta-stable vacua. 
As we mentioned in section \ref{sec:msv}, the vacuum has SU($N$)$_{c+f}$
color-flavor locking symmetry.
This symmetry is broken down to U(1)$\times$SU($N-1$) by the single
non-Abelian semilocal string (\ref{embedding}), so that 
the solution has internal orientational zero modes (Nambu-Goldstone modes)
${\bf C}P^{N-1} \simeq SU(N)/ [SU(N-1)\times U(1)]$.

When embedding the U(1)
semilocal solution to the non-Abelian gauge theory, 
we have to choose a single U(1) gauge sub-sector
in the U($N$). This corresponds precisely to choosing a single 
D4-brane (among $N=N_f-N_c$ of them) on which the D2-brane should end.
The tension of the non-Abelian semilocal string is proportional to 
$\mu^2$. 
This is again consistent with 
the D-brane picture, where the length of
the D2-brane (proportional to $\mu^2$) is determined only by the
distance between the NS5-brane and the $(N_f-N_c)$ D4-branes  
and independent of the other parameters.

Now we consider a special case where the Higgs self-coupling $h$
satisfies
\begin{eqnarray}
 g^2 = |h|^2,
\label{g=h}
\end{eqnarray}
although this would not be always satisfied for the dual of SQCD.
In this csae the non-Abelian semilocal string saturates 
the Bogomol'nyi bound of the theory with
(\ref{eqsemi})  (the tension is given by $4\pi \mu^2$), though 
the supersymmetries are broken because of the cosmological constant.
In this case the second order differential equation (\ref{2ndeom}) 
reduces to the 1st order differential equation (\ref{bpseqano})
with replacing $v^2$ with $\mu^2$.
It is known that the minimal BPS semilocal vortex
has the size moduli in addition to the orientational moduli ${\bf C}P^{N-1}$. 
Furthermore, any repulsive or attractive force does not appear
between separated strings due to the saturation, like the
ordinary 1/2 BPS solitons. As a result the moduli space of the non-Abelian
semilocal string in the meta-stable SUSY breaking vacua is completely 
the same as that for the 1/2 BPS non-Abelian semilocal string in 
${\cal N}=2$ supersymmetric vacua
\cite{Hanany:2004ea,Auzzi:2003fs,Auzzi:2003em,mmf}. 

We would like to stress that our non-Abelian semilocal string solutions
are always non-BPS (breaking all the supersymmetries) 
even for the case where
the energy bound is saturated ($g^2 = |h^2|$). 
One can easily show that
the 1/2 BPS equation (\ref{bpseqano}) is inconsistent with
the ${\cal N}=1$ super-transformation of the gaugino 
$\delta_\epsilon \lambda = \sigma^{\mu\nu}\epsilon F_{\mu\nu} 
+ i\epsilon D$
with $D \sim qq^\dagger - \tilde q^\dagger \tilde q = 0$ in the
selected sector (\ref{assumption}).
It is somehow surprising that 
the supersymmetry-breaking semilocal
string saturates the 
Bogomol'nyi energy bound 
when $g^2 = |h|^2$ and
the non-BPS semilocal strings behaves as if they were 
1/2 BPS
semilocal strings.\footnote{
Our non-BPS string 
is not a kind of the $F$-term vortex string in ${\cal N}=1$ theory 
\cite{Davis:1997bs,Achucarro:2004ry} 
which also behaves as if it was a 1/2 BPS string though it doesn't 
preserve the ${\cal N}=1$ supersymmetry. 
The $F$-term string preserves 1/2 supersymmetry
when it is embedded into an appropriate ${\cal N}=2$ theory.
However, our non-BPS strings cannot be simply embedded into any 
${\cal N}=2$
model, partly because the scalar field $M$ is not an adjoint scalar 
field of the $U(N)$.  
This may be also obvious from the viewpoint of the D-brane configuration,
see Fig.\ref{dht_n1}.} 
Actually they have free moduli parameters of
positions, internal orientations and sizes. From the D-branes viewpoint
this is not intuitive at all since the angle between
the D2-branes and D4'-branes is  not the right angle.

Due to the saturation in the case where $g^2 = |h|^2$, 
the string is stable in the selected sector
(\ref{assumption}).  
Later we will study the fluctuation orthogonal to the truncation
(\ref{assumption}), and find that the string is stable against all the
fluctuations when (\ref{g=h}) is satisfied.

\subsection{Stability of solitonic string}

Let us study the stability of the ANO string in the magnetic theory of
the massive SQCD, which we found in the previous subsection.
We will find that the vortex is stable for 
\begin{eqnarray}
 g^2 \geq |h|^2.
\label{stability}
\end{eqnarray}

It is known \cite{Vachaspati:1991dz} that Abelian semilocal vortices are
unstable for 
$g^2<|h|^2$. In this parameter region the strings can reduce their
tension by fattening themselves (increasing their widths). 
Therefore, any possible region of the parameter space for the ANO string 
to be stable should be within (\ref{stability}). In our truncated theory
with (\ref{assumption}), this Abelian situation corresponds to
$N=1$. For our general non-Abelian case, we can show that the
non-Abelian semilocal string is stable again for
(\ref{stability}), as in the following. The solution
(\ref{embedding}) for $g^2>|h|^2$ has vanishing 
$\{\phi_1^\star, \cdots, \phi_{N_f-N}^\star\}$, thus  we consider a fluctuation
from this ANO solution. Matrix elements of the scalar fluctuation are
labeled as 
\begin{eqnarray}
\phi =  \left(
\begin{array}{ccc}
\sqrt{2}\mu {\bf 1}_{N-1}+ \delta \phi_{I} & \delta \phi_{II} & \delta \phi_{III}\\
\delta \phi_{IV} & \phi_0^*+ \delta\phi_{V} & \delta \phi_{VI}
\end{array}
\right)
\end{eqnarray}
Substituting this and a similar decomposition of the gauge fields into
the Lagrangian, it is easy to show that the ``semilocal sectors''
$\delta\phi_{III}$ and $\delta\phi_{VI}$ decouple from 
the others (which are fluctuations of a non-Abelian local string with
$N=N_f$).  
This latter fluctuations are expected to be stable for any $g$ and $h$.
$\delta\phi_V$ is found to be identical to the fluctuation of the Abelian
local vortex (ANO), thus is stable. The sector $\delta\phi_I$ couples to
the vacuum expectation value $\sqrt{2}\mu$ and is Higgsed to be massive
and stable. A combination of $\delta\phi_{II}$ and $\delta\phi_{IV}$
produces the orientational moduli which are normalizable, localized and
massless modes. The other combinations are massive and stable, through a
Higgs mechanism, which would
be understood in a unitary gauge, for example.
Our main concern is in the semilocal sector. We find that
$\delta\phi_{III}$ does not have a potential term and thus
is a massless bulk mode which is non-normalizable. 
This is a Nambu-Goldstone mode associated with the global symmetry
breaking by the vacuum.
$\delta\phi_{VI}$
has the same expression as the fluctuation of the Abelian semilocal
string, therefore we know that for $g^2>|h|^2$ it is massive and stable, as
mentioned above. For $g^2=|h|^2$, this $\delta \phi_{VI}$ gives the massless moduli
$\{\delta \phi_1^\star, \cdots, \delta \phi_{N_f-N}^\star \}$
as indicated in (\ref{embedding}).
In sum, the solution (\ref{embedding}) is stable for
(\ref{stability}) in the truncated model (\ref{assumption}).

Second, we consider fluctuations orthogonal to the truncation
(\ref{assumption}). Fluctuation of the field $M$ does not provide any
instability, because in the potential term (\ref{iss_pot_1})
the field $M$ appears in a positive-semi-definite form. Thus we need
to study 
the fluctuation of $q$ and $\widetilde{q}$. 
We redefine the fields as
\begin{eqnarray}
 \phi \equiv \frac{1}{\sqrt{2}}(q + \widetilde{q}^\dagger),
\quad 
 \widetilde{\phi} 
\equiv \frac{1}{\sqrt{2}}(q - \widetilde{q}^\dagger).
\end{eqnarray}
(Previously we have put $\widetilde{\phi}=0$.) The $\widetilde{\phi}$
potential can be written, up to its quadratic order, as 
\begin{eqnarray}
 {\rm Tr}_f\!
\left[
|\mu^2 h^2| \widetilde{\phi}^\dagger \widetilde{\phi}
+ \frac{|h|^2}{2}\phi^\dagger \widetilde{\phi} 
\widetilde{\phi}^\dagger \phi\right]
+ \frac14 (g^2\!-\!|h|^2)\;
 {\rm Tr}_c\!
\left[\left(
\widetilde{\phi} \phi^\dagger + \phi \widetilde{\phi}^\dagger
\right)^2
\right].
\label{phitildepo}
\end{eqnarray}
The term linear in $\widetilde{\phi}$ vanishes identically, and
consequently there is no mixing between $\widetilde{\phi}$ and 
the fluctuation of $\phi$.
Since (\ref{phitildepo}) is a sum of perfect squareds, 
the fluctuation is stable when (\ref{stability}) is
satisfied. Therefore we conclude that the ANO string is stable for 
$g^2\geq |h|^2$.

Irrespective of the values of the couplings, there exist massless
fluctuations. The eigen function of the fluctuation
is given for example by 
\begin{eqnarray}
\delta \phi_{(i,j)} =  \phi^{\rm ANO}(x^1,x^2)f_{(i,j)}(x^0,x^3)
\end{eqnarray}
where $(i,j)$ labels the matrix elements, $x^3$ is the direction along
the embedded ANO string, 
and $\phi^{\rm ANO}$ is the ANO vortex solution in the Abelian Higgs
model. 
This is
certainly expected, since this is a Nambu-Goldstone mode associated with
the spontaneous breaking of the global symmetry SU($N_f$). If we rotate
the embedded ANO solution a little bit by the global symmetry SU($N_f$),
then we obtain this massless fluctuation.

\subsection{Multi-tension non-BPS vortex strings}
\label{sec:multi}

Instead of choosing all the ``mass'' equal to each other (to be equal to
$\mu^2$), we may choose a superpotential 
$W = -h{\rm Tr}_f\left[mM\right]$, as described in section
\ref{sec:iss}.  
We align the mass eigenvalues as 
$m = {\rm diag} (m_1,\cdots,m_{N_f})$ with
$m_1\geq \cdots\geq m_{N_f}\geq 0$ so that the 
vacuum is stable locally (see \cite{OO2} for discussions on the vacuum
stability from the brane configurations). 

Obviously we find the
embedding of the ANO string as before,
\begin{eqnarray}
\phi_{(1,1)} = \phi^{\rm ANO}(x^1,x^2), \quad 
A_{(1,1)}^\mu = (A^{\rm ANO})^\mu(x^1,x^2), \quad 
{\rm the \;\; others}=0.
\label{embed11}
\end{eqnarray}
Here the subscript means again the $(i,j)$ component. 
For $g^2=|h|^2$, the tension of this ANO string is proportional to
$m_1$, as
predicted from the brane configurations. 

One can embed the string in one of the other sectors, by replacing
the above $(1,1)$ by $(j,j)$ for $j=2,\cdots, N_f-N_c$. This is again
obviously a solution to the equations of motion. The ANO string has its
tension $\propto m_i$, which is consistent with the brane
configurations. Thus we actually obtain multi-tension vortices.

The analysis of the stability of the string is almost similar
to the case of the semilocal strings. For example, if we embed an
ANO string in the $(1,1)$ component as in (\ref{embed11}), we find
the fluctuation Lagrangian for $\phi_{(j,1)}$ for $j=2,\cdots, N_f$
as
\begin{eqnarray}
- |{\cal D}_\mu^{\rm ANO}\phi_{(j,1)}|^2 - \frac{|h|^2}{2}
\left(
|\phi^{\rm ANO}|^2 -2m_{j}
\right)|\phi_{(j,1)}|^2 .
\label{flucmnf}
\end{eqnarray}
For the semilocal case 
$m_1=\cdots=m_{N_f}=\mu^2$, we know that this fluctuation is stable for
$g^2\geq |h|^2$. In the present case, the above fluctuation 
Lagrangian is obtained just by replacing $\mu^2$ by $m_j$. 
Since $m_j$ is smaller than $m_1$, the tachyonic instability 
is now found to be improved, and the situation is better for the
stability.  
Thus for $g^2\geq |h|^2$ the ANO solution
is stable against this kind of fluctuations.
For the fluctuations orthogonal to the truncation (\ref{assumption}), 
a similar computation shows that the potential is the same as 
(\ref{phitildepo}) except that we replace $|\mu|^2$ in the first term by
the 
matrix $m$. The argument for the stability is the same, and we find that 
the string is stable for $g^2\geq |h|^2$ against $\tilde{\phi}$. 
Therefore, in summary, we find that the ANO string embedded in $(1,1)$
sector is stable for $g^2\geq |h|^2$.

When $m_{N+1}=\cdots = m_{N_f}=0$, the vacuum admits supersymmetries, 
and the cosmological constant vanishes. 
In the corresponding brane configuration, 
all the D4-branes become parallel, which is a manifestation of the
supersymmetry restoration. Let us consider the embedding of the ANO
string as before, in this supersymmetric vacuum. The embedded 
solitonic string breaks the supersymmetries
 even for $g^2 = |h|^2$,
similarly to the case of the non-Abelian semilocal strings
dealt with at the end of section \ref{sec:classical sol}.
This is consistent with what we have found in the brane configurations.

\section{Conclusion and Discussions}
\label{sec:conclusion}

The vortex strings which we have found in this paper can be applied to
various situations, such as phenomenological model building and
cosmologies. Since the meta-stable vacua found in \cite{ISS} provide
us with a new path to break supersymmetries at low energy, the possible
existence of solitons in the vacua may affect any story on the vacua.
We have found that the $U(N_c)$ and the $SO(N)$ SQCD
have vortex strings. The vortex strings in the $U(N_c)$ (and the
$SO(N_c)$ with $N_f=N_c-2$) are similar to the semilocal
strings. The vortex strings in general $SO(N_c)$ are ${\bf Z}_2$ strings.
The existence and the properties of the $U(N_c)$ 
strings have been found
from brane configurations. This time again, stringy technique turned out
to be quite useful in finding field theoretical solutions and their
properties. 

Several discussions and comments are in order, which we hope to 
get back to in the future.
\begin{itemize}
 \item Implication to cosmologies. The vortex strings found in this
       paper can be thought of as cosmic strings. One can argue that
when the universe is cooling down, the energy scale gets smaller than
       the typical scale determined by the Seiberg duality, and if
       eventually the supersymmetry-breaking meta-stable vacua are
       chosen somehow, the vortex strings may form then. It is
       noteworthy that, even though the vacua is non-supersymmetric and
       consequently the cosmic string is non-BPS, when 
the gauge coupling $g$ is equal to the scalar self coupling $|h|$, the
       solution itself is the same as the BPS vortex strings. Therefore
       one can use various results on the BPS vortex strings found 
       in particular in the moduli matrix formalism \cite{mmf} (see 
       \cite{inprogress}). For example, 
       reconnection probability of cosmic strings is important for
       evaluating number density of the cosmic strings and consequently 
       possible observation of them. The cosmic strings found in this
       paper with $g=|h|$ provides an analytic study of the reconnection
       of cosmic strings in a non-supersymmetric background, which
       is quite interesting.
\item Renormalization group flow and stability of the strings. The
       stability analysis of section \ref{sec:solution} has been done at
       the tree 
       level of the magnetic theory. There we have found that for 
       $g\geq |h|$ the string is stable. To argue the stability in 
       more realistic situations, one needs to consider quantum
       effects, including the renormalization group flow of the
       couplings. It is important to know if the stability condition
       $g\geq |h|$ is satisfied or not at the energy scale where
       the classical soliton solutions are reliable. The
       stability depends on the beta functions and the precise values of
       the coupling constants at some energy scale. However, it is
       difficult to determine the Yukawa coupling $h$ precisely, though
       we know it is ${\cal O}(1)$ at the typical energy scale
       appearing in the Seiberg duality.
\item Prediction of the stability from brane configurations.
As studied in section \ref{sec:solution}, the vortex string is
       classically stable for $g\geq|h|$. It is interesting if this
       condition can be seen in the brane configurations.
       The couplings $g$ and $h$ can be interpreted \cite{Bena:2006rg}
       as the distances 
       between D-branes and NS5(NS'5)-branes (at least for the massless
       SQCD), as shown in 
       Fig.\ref{dht_n1_2}. The distance between NS5-NS'5 along the $x^6$
       axis corresponds to the gauge coupling as $\sim 1/g^2$, that
       between NS5-D6 along the $x^6$ axis to $\sim 1/|h|^2$ and NS5-D6
       along the $x^4$ axis to the ``mass'' $\sim \mu^2$. 
\begin{figure}[t]
\begin{center}
\includegraphics[width=13cm]{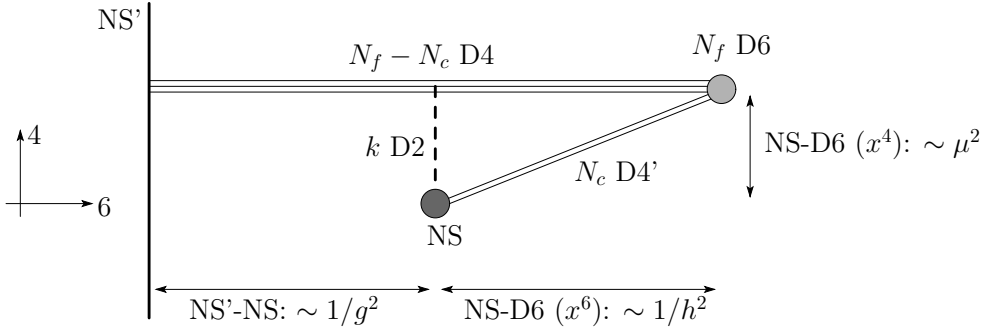}
\caption{The relation between the field theory couplings and locations
 of the branes.}
\label{dht_n1_2}
\end{center}
\end{figure}
       When the distance between NS5-NS'5 is smaller than the distance
       NS5-D6 along the $x^6$ axis, we have $g^2 > |h|^2$, so the field
       theory results shows that the ANO string (the semilocal string
       with vanishing width) stably exists there. When the distance
       between NS5-NS'5 is smaller than the distance NS5-D6 along the
       $x^6$ axis, we have $g^2 < |h|^2$ so any string solution does
       not exist in field theory. 
The question is if one can read these stability information purely in the
       brane configuration. Our tentative conclusion here is that we can
       read the tendency of the instability, but the actual relation
       $g\geq |h|$ is difficult to be seen, as is explained below.

The moduli space of the solitonic string should be seen along the argument
given by Hanany and Tong \cite{Hanany:2004ea}, as a low energy field
       theory on the D2-brane. 
As seen in section \ref{sec:rev}, for the semilocal strings in the
       ${\cal N}=2$ gauge theory, massless excitations on the D2-brane
       came from strings  
in the following three sectors:
(i) $Z$ from D2-D2, (ii) $\psi$ from D2-D4 and 
(iii) $\tilde \psi$ from D4'-D2 in Fig.\ref{ht}. (i) is relevant for the 
transverse location of the string, (ii) is for the orientational moduli,
       and (iii) is for the width of the string. Apparently the
       instability of the semilocal string should be related to (iii).

Suppose that NS5-brane is closer to the D6-branes, compared to the
       distance to the NS'5-brane. 
As can be found in Fig.\ref{dht_n1} or in Fig.\ref{dht_n1_2}, 
the D2-D4' string is naively expected to become tachyonic, since the
       angle between the D2-brane and the D4'-brane is not the right
       angle. Consequently, the solitonic string is expected to be
       unstable against the change of the width. This is consistent with
       the field theory analysis for $g<|h|$. When one moves the
       NS5-brane toward the NS'5-brane in the $x^6$ direction, the angle
       becomes closer to the right angle, thus the tachyonic 
       instability is improved. This is again consistent with the field
       theory analysis, since $g/|h|$ is getting larger.

However, a strange discrepancy between the D-brane picture and the field
       theory analysis appears when one continues to make the NS5-brane
       approach the NS'5-brane. The angle is still less than the right
       angle and thus the D2-D4' string seems to be tachyonic, while in
       the field theory analysis the vortex becomes stable. This tells
       us that the angle is not directly corresponding to the mass 
       of the moduli parameter, somehow, although in the ${\cal N}=2$
       case \cite{Hanany:2004ea} this has been assumed even in the
       presence of 
       the NS5-brane and it worked. Our situation breaks
       supersymmetries, that might be a reason why it does not work now.

A similar argument can apply for the case of multi-tension strings found
       in section \ref{sec:multi}. We have found in (\ref{flucmnf}) that
the string is more stable for smaller $m_j$. 
This phenomenon is compatible with the brane configuration. Now the
D2-brane is ending on the D4-brane whose position in the $x^4$-$x^5$
plane is given by the complex parameter $m_1$. When other mass
parameters $|m_i|$ ($i=N+1,\cdots, N_f$) are smaller than this $|m_1|$, 
the angle between the D2-brane and one of 
the D4-branes labeled by $i=N+1,\cdots,N_f$ is closer to the right
angle, thus the instability of the D2-brane caused by the tachyonic 
strings connecting the D2-brane and the D4-branes is smaller. 
However, again the problem of vanishing instability still remains in the
       brane story.
\end{itemize}

\acknowledgments 

 We are grateful to K.~Intriligator for his kind correspondence
and comments.
K.H. would like to thank T.~Hirayama for useful discussions.
M.E.~would like to thank the theoretical HEP group of KIAS.
The work of M.E.~is supported by Japan Society for the Promotion 
of Science under the Post-doctoral Research Program. 
K.H.~is partly supported by
the Japan Ministry of Education, Culture, Sports, Science and
Technology. K.H.~would like to thank 
the Yukawa Institute for Theoretical Physics at Kyoto University, where
this work was discussed during the workshop YITP-W-06-11 on ``String
Theory and Quantum Field Theory''.


\appendix

\section{ANO like vortex in the $SU(N) \times U(1)_B$ gauge theory}

In this Appendix, 
we construct ANO like vortex in the 
magnetic theory where the $SU(N)$ gauge coupling $g$ and 
the $U(1)_B$ gauge coupling $e$ are different.
Here we consider the field configuration (\ref{assumption}) only
and then the the equations of motion is
\begin{eqnarray}
{\cal D}_\mu {\cal D}^\mu \phi &=& \frac{\partial \widetilde V}{\partial \phi^\dagger}, \\
\frac{1}{g^2} \left[ {\cal D}^\mu F_{\mu\nu} \right]^{SU(N)} &=& 
-\frac{i}{2}\left[ \left(\phi {\cal D}_\nu \phi^\dagger
- {\cal D}_\nu \phi \phi^\dagger \right) \right]^{SU(N)},  \\
\frac{1}{e^2} \left[ {\cal D}^\mu F_{\mu\nu} \right]^{U(1)} &=& 
-\frac{i}{2}\left[ \left(\phi {\cal D}_\nu \phi^\dagger
- {\cal D}_\nu \phi \phi^\dagger \right) \right]^{U(1)}, 
\label{2ndeom2}
\end{eqnarray}
where $[\cdots ]^{SU(N)}$ and $[\cdots ]^{U(1)}$ 
means the traceless part and the trace part
of the $N \times N$ matrix, respectively.
It is easy to see that 
the solution of the equations of motion in 
this restricted configuration space
is also the solution in the full configuration space.
Here we take the following axial ansatz for the solution (see 
\cite{Auzzi:2003fs,Shifman:2006kd}):
\begin{eqnarray}
A_2 - i A_1 = \frac{n}{r}
\left(
\begin{array}{cccc}
b(r) & & & \\
& \ddots & & \\
& & b(r) & \\
& & & a(r)
\end{array}
\right)e^{i\theta},\ 
\phi = 
\left(
\begin{array}{ccccccc}
v(r) & & & & & & \\
& \ddots & & & & & \\
& & v(r) & & \\
& & & u(r) e^{i n \theta}
& 0 & \cdots & 0
\end{array}
\right),
\label{embedding2}
\end{eqnarray}
where $n$ is an integer
and all the other components of $A_\mu$ are zero.
Then the equations of motion become
\begin{eqnarray}
&& u''+ \frac{u'}{r}-n^2\frac{(1-a)^2}{r^2}u 
=\frac{|h|^2}{2}(u^2 - 2 \mu^2) u, 
\nonumber \\
&& v''+\frac{v'}{r}-n^2\frac{b^2}{r^2}v 
=\frac{|h|^2}{2}(v^2 - 2 \mu^2) v, 
\nonumber \\
&& \left( \frac{N-1}{g^2} +\frac{1}{e^2} \right)
\left( a''-\frac{a'}{r} \right)+
\left( \frac{1}{e^2} -\frac{1}{g^2} \right) (N-1)
\left( b''-\frac{b'}{r} \right) = - N (1-a) u^2, 
\nonumber \\
&& \left( \frac{1}{g^2} +\frac{N-1}{e^2} \right)
\left( b''-\frac{b'}{r} \right)+
\left( \frac{1}{e^2} -\frac{1}{g^2} \right)
\left( a''-\frac{a'}{r} \right) = N b v^2, 
\label{4eq}
\end{eqnarray}
where $a'=\frac{d}{dr}a(r)$ and so on.
The boundary conditions at the origin should be 
\beq
a(0)=b(0)=u(0)=v'(0)=0
\eeq
because $\phi, \partial_\mu \phi$ and $F_{\mu \nu}$ 
are regular at the origin.
Actually, 
\beq
a(r) ={\cal O} (r^2), b(r) ={\cal O} (r^2),
u(r) ={\cal O} (r^n), v'(r) ={\cal O} (r^1),
\eeq
are consistent with (\ref{4eq}).
The boundary conditions at $r=\infty$ should be 
\beq
a(\infty)=1, \;
b(\infty)=0, \;
u(\infty)=v(\infty)=\sqrt{2}\mu,
\eeq
which is also consistent with (\ref{4eq}).
Then, 
with these eight consistent boundary conditions
we can solve (\ref{4eq}) in principle
as in the case of the simple ANO vortex
though we do not carry out it here.
Therefore we expect 
(\ref{embedding2}) is the vortex solution with $n$ charge.
Note that 
\begin{eqnarray}
\pi_1
\left( \frac{ SU(N) \times U(1)_B }{{\bf Z}_N} \right)
= {\bf Z}
\end{eqnarray}
and the vortex with the minimal charge winds 
the $U(1)_B$ $1/N$ times.

Finally, we comment on the stability of the solutions.
If $|e|>|h|$ these solutions is expected to be stable
as in the case $e=g$.
A more complete analysis of the stability is desired,
but, we will leave it as a future problem.

\newcommand{\J}[4]{{\it #1} {\bf #2} (#3) #4}
\newcommand{\andJ}[3]{{\bf #1} (#2) #3}
\newcommand{\AP}{Ann.\ Phys.\ (N.Y.)}
\newcommand{\MPL}{Mod.\ Phys.\ Lett.}
\newcommand{\NP}{Nucl.\ Phys.}
\newcommand{\PL}{Phys.\ Lett.}
\newcommand{\PR}{ Phys.\ Rev.}
\newcommand{\PRL}{Phys.\ Rev.\ Lett.}
\newcommand{\PTP}{Prog.\ Theor.\ Phys.}
\newcommand{\hep}[1]{{\tt hep-th/{#1}}}

\end{document}